\documentclass[10pt,twocolumn,letterpaper]{article}

\usepackage[pagenumbers]{cvpr} % To force page numbers, e.g. for an arXiv version

\usepackage{graphicx}
\usepackage{amsmath}
\usepackage{amssymb}
\usepackage{booktabs}
\usepackage{makecell}
\usepackage{multicol}
\usepackage{multirow}
\usepackage{eurosym}
\usepackage{comment}
\usepackage{bbding,pifont}
\usepackage{tabularx}
\newcommand{\cmark}{\ding{51}}%
\newcommand{\xmark}{\ding{55}}%

\usepackage[pagebackref,breaklinks,colorlinks]{hyperref}
\usepackage[accsupp]{axessibility}

\usepackage[capitalize]{cleveref}
\crefname{section}{Sec.}{Secs.}
\Crefname{section}{Section}{Sections}
\Crefname{table}{Table}{Tables}
\crefname{table}{Tab.}{Tabs.}

\def\cvprPaperID{3945} % *** Enter the CVPR Paper ID here
\def\confName{CVPR}
\def\confYear{2023}

\begin{document}

\title{A Simple Baseline for Video Restoration with Grouped Spatial-temporal Shift}

\author{
Dasong Li$^{1}$ \and
Xiaoyu Shi$^{1}$ \and
Yi Zhang$^{1}$ \and
Ka Chun Cheung$^{2}$ \and
Simon See$^{2}$ \and
Xiaogang Wang$^{1,4}$ \and
Hongwei Qin$^{3}$ \and
Hongsheng Li$^{1,4}$ \\ \and
$^{1}$CUHK MMLab \quad
$^{2}$NVIDIA AI Technology Center \quad $^{3}$SenseTime Research \quad
$^{4}$CPII under InnoHK \\
{\tt\small \{dasongli@link, hsli@ee\}.cuhk.edu.hk}
}
\maketitle

%%%%%%%%% ABSTRACT
\begin{abstract}
 Video restoration, which aims to restore clear frames from degraded videos, has numerous important applications. The key to video restoration depends on utilizing inter-frame information. However, existing deep learning methods often rely on complicated network architectures, such as optical flow estimation, deformable convolution, and cross-frame self-attention layers, resulting in high computational costs. In this study, we propose a simple yet effective framework for video restoration. Our approach is based on grouped spatial-temporal shift, which is a lightweight and straightforward technique that can implicitly capture inter-frame correspondences for multi-frame aggregation. By introducing grouped spatial shift, we attain expansive effective receptive fields. Combined with basic 2D convolution, this simple framework can effectively aggregate inter-frame information. Extensive experiments demonstrate that our framework outperforms the previous state-of-the-art method, while using less than a quarter of its computational cost, on both video deblurring and video denoising tasks. These results indicate the potential for our approach to significantly reduce computational overhead while maintaining high-quality results. Code is avaliable at \url{https://github.com/dasongli1/Shift-Net}.
\end{abstract}

\section{Introduction}

The popularity of capturing videos using handheld devices continues to surge. However, these videos often suffer from various types of degradation, including image noise due to low-cost sensors and severe blurs resulting from camera shake or object movement. Consequently, video restoration has garnered significant attention in recent years.

The keys of video restoration methods lie in designing components to realize alignment across frames. While several methods~\cite{EMVD,fastdvdnet,RNN-MBP,MildenhallKPN18,BPN} employ convolutional networks for multi-frame fusion without explicit alignment, their performance tends to be suboptimal.
Most methods rely on explicit alignment to establish temporal correspondences, using techniques such as optical flow ~\cite{toflow,shi2023videoflow} or deformable convolution~\cite{dai2017deformable,zhu2019deformable}. 
However, these approaches often necessitate either complex or computationally expensive network architectures to achieve large receptive fields,  and they may fail in scenarios involving large displacements \cite{ARVo}, frame noise \cite{SeeMotion,RViDeNet}, and blurry regions \cite{Son2021PVDNet,RNN-MBP}.
Recently, transformer ~\cite{devlin2018bert,dosovitskiy2021an,liu2021Swin} becomes promising alternatives for attaining long-range receptive fields. A video restoration transformer (VRT) \cite{liang2022vrt} is developed to model long-range dependency, but its large number of self-attention layers make it computationally demanding. Inspired by the success of the Swin transformer \cite{liu2021Swin}, large kernel convolutions \cite{convNet2020, replknet} emerge as a direct solution to obtain large effective receptive fields. 
However, extremely large kernels (e.g. kernel size $>$ 13$\times$13) cannot necessarily guarantee improved performance (shown in \ref{ab_largeKernel}).
\begin{figure}[!t]
    \begin{center}
    \includegraphics[width=.87\linewidth]{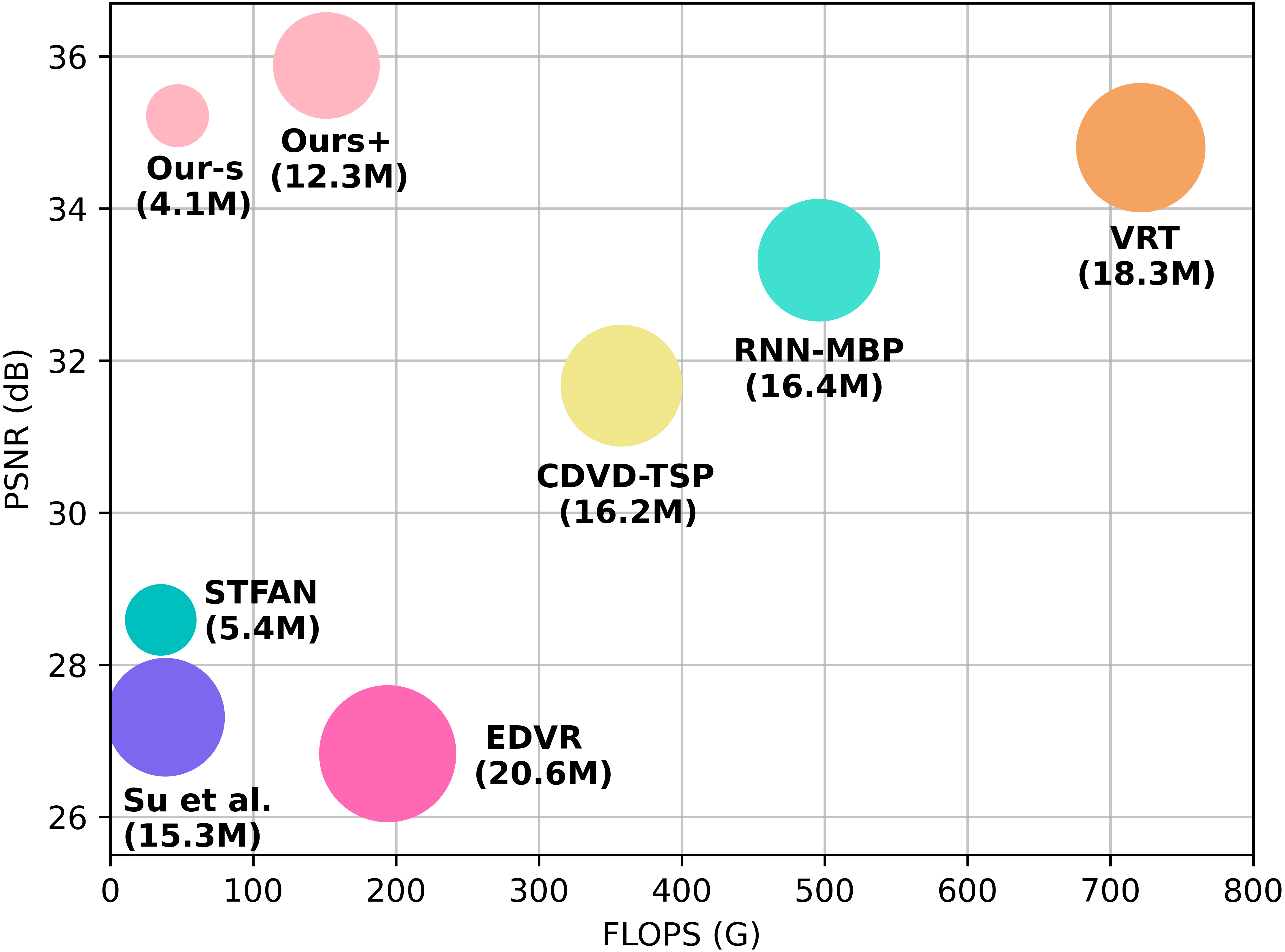}
    \end{center}
    \vspace{-0.6cm}
    \caption{Video deblurring on GoPro dataset \cite{deblur-multi-scale}. 
    Our models have fewer parameters (disk sizes) and occupy the top-left corner, indicating superior performances (PSNR on y-axis) with less computational cost (FLOPS on x-axis). 
    }
    \label{fig:vis_flops_psnr}
    \vspace{-0.6cm}
\end{figure}

\begin{figure*}
    \centering
    \includegraphics[width=0.88\textwidth]{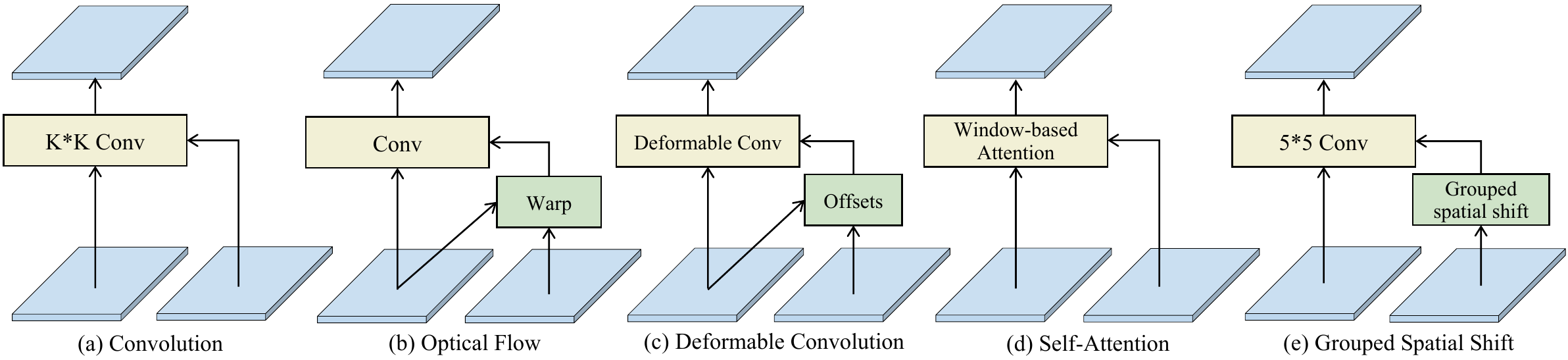}
    \vspace{-0.3cm}
    \caption{Different modules for multi-frame aggregation: a) convolution \cite{fastdvdnet}, b) optical flow \cite{Pan2020cdvdtsp,liang2022vrt}, c) deformable convolution \cite{dai2017deformable,tian2020tdan,wang2019edvr}, d) self-attention \cite{liu2021Swin,liang2022vrt} and e) our grouped spatial shift. Point-wise convolution, shortcut and normalization are omitted for simplicity.
    } 
    \label{fig:brief_pic}
    \vspace{-0.4cm}
\end{figure*}

In this study, we propose a simple, yet effective spatial-temporal shift block to achieve large effective receptive field for temporal correspondence.
We introduce a Group Shift-Net, which incorporates the proposed spatial-temporal shift blocks for alignment along with basic 2D U-Nets for frame-wise feature encoding and restoration. The grouped spatial-temporal shift process involves the separate shifting of input clip features in both temporal and spatial dimensions, followed by fusion using 2D convolution blocks. Despite its minimal computational demands, the shift block offers large receptive fields for efficient multi-frame fusion.
By stacking multiple spatial-temporal shift blocks, the aggregation of long-term information is achieved. This streamlined framework models long-term dependencies without depending on resource-demanding optical flow estimation \cite{toflow,huang2022flowformer,shi2023flowformer++}, deformable convolution \cite{dai2017deformable,tian2020tdan,wang2019edvr}, 
or self-attention \cite{liang2022vrt}.

Notably, while temporal shift module (TSM) \cite{lin2019tsm} was originally proposed for video understanding, it is not effective for video restoration. Our method distinguishes itself from TSM in three fundamental ways: a) \textit{Alternative bi-directional temporal shift.} TSM \cite{lin2019tsm} employs bi-directional \textit{channel} shift during training, causing misalignment of channels across three frames, which in turn increases the difficulty of multi-frame aggregation. Conversely, our method utilizes alternative \textit{temporal shifts}, effectively circumventing this issue. b) \textit{Spatial shift.} In addition, our approach also incorporates a spatial shift for multi-frame features. We divide the features into several groups, each with distinct shift lengths and directions in the 2D dimension. This grouped spatial shift offers multiple candidate displacements for matching misaligned features. c) \textit{Feature fusion.} To seamlessly merge various shifted groups, the kernel size of the convolution is set equal to the base shift length. By combining elements b) and c), the spatial-temporal shift achieves large receptive fields (e.g. $23\times 23$).

The contributions of this study are two-fold: 1) We propose a simple, yet effective framework for video restoration, which introduces a grouped spatial-temporal shift for efficient and effective temporal feature aggregation 2) Our framework surpasses state-of-the-art methods with much fever FLOPs on both video deblurring and video denoising tasks, demonstrating its generalization capability.

\begin{figure*}
    \centering
    \includegraphics[width=0.78\textwidth]{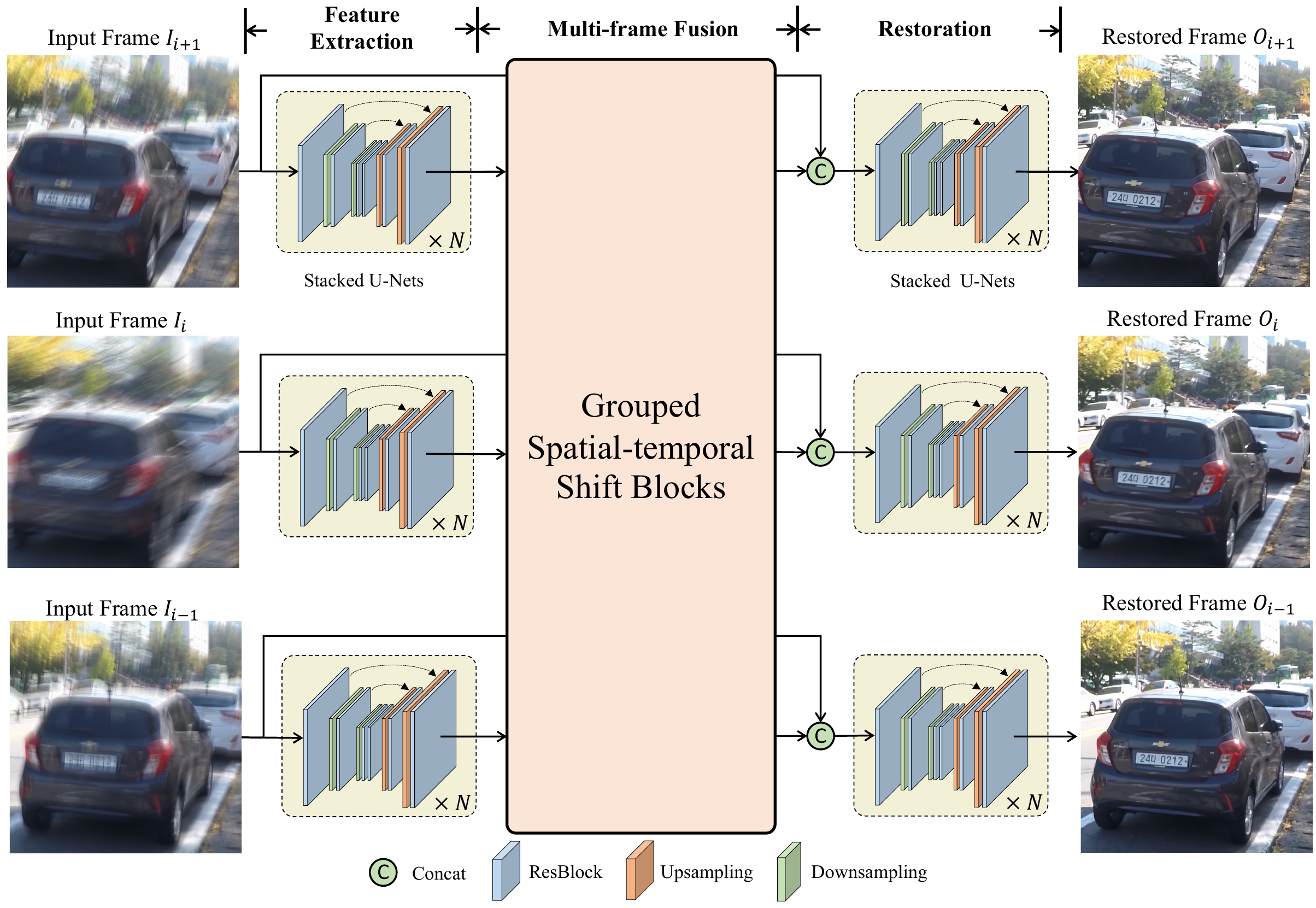}
    \vspace{-0.4cm}
    \caption{Overview of the Group Shift-Net. It adopts a three-stage design: feature extraction, multi-frame fusion, and final restoration. Grouped spatial-temporal shift blocks are proposed to achieve multi-frame aggregation.
    } 
    \label{fig:overview}
    \vspace{-0.4cm}
\end{figure*}

\section{Related Work}
A series of methods have been proposed to explore \textit{temporal information} for video restoration.
 
\noindent\textbf{Temporal alignment.} 
Temporal alignment is a vital step to model temporal correspondences of misaligned frames in videos.
Early learning-based methods \cite{VSRCNN,real-timeVSR,dvdnet,Son2021PVDNet,Li2022efficient} employ traditional image alignment methods \cite{deepflow} to model the motions.
To handle complicated motions, Xue et al. \cite{toflow} propose task-oriented flow by fine-tuning a pretrained optical flow model \cite{Spynet} on different video restoration tasks.  
Dynamic filters \cite{Jo2018dynamic_upsampling,zhou2019stfan} are also proposed to achieve motion compensation. Tian et al. \cite{tian2020tdan, wang2019edvr} propose to utilize deformable convolution \cite{dai2017deformable} for feature alignment. 
Chan et al. \cite{chan2021basicvsr++} leverage the optical flow to guide the deformable alignment for stable training \cite{chan2021understanding}, which is also adopt by the latest transformer-based method VRT \cite{liang2022vrt}. Such alignment techniques increase the model complexity and might fail in the case of large displacement \cite{ARVo}, noise \cite{SeeMotion,RViDeNet,zhang2023kbnet}, blurry regions \cite{Son2021PVDNet,RNN-MBP}. Zhu et al. \cite{RNN-MBP} demonstates that optical flow or deformable convolution cannot estimate the alignment information well because of the significant influence of the motion blur. 
A series of methods \cite{fastdvdnet,RNN-MBP,EMVD} are proposed to utilize convolution networks to handle motion implicitly. However, the networks with small kernel sizes usually have narrow receptive fields \cite{luo2017understanding}, which limits the model capacity to address large displacements. 
 
\noindent\textbf{Long-term information aggregation.} To obtain the long-term information from distant frames, learning-based methods can be classified as sliding window-based methods and recurrent methods. Sliding window-based methods \cite{fastdvdnet,Pan2020cdvdtsp} usually take several adjacent frames as input and output the center restored frame. The information can only be aggregated within the fixed sliding window. 
In contrast, several methods \cite{RSDN,Son2021PVDNet,EMVD,chan2021basicvsr,chan2021basicvsr++} utilize the recurrent framework for long-term information aggregation. The faulty prediction and misalignment are accumulated frame by frame, which may deteriorate the long-term dependency modeling \cite{chan2022investigating}. 

\noindent\textbf{Shift operations.} Wu et al. \cite{wu2018shiftconv} combine shift operation and $1\times1$ convolution as an efficient alternative to $3\times3$ convolution. Its variants \cite{Shift_conv_fast,shifts_conv_need} further propose learnable active shifts.
Zhang et al. \cite{zhang2022efficient} adopt shift and $1\times1$ convolution for efficient image super-resolution.
Lin et al. \cite{lin2019tsm} propose a temporal shift module (TSM) for video understanding. Rong et al. \cite{tsmdenoising} apply temporal shift on wavelet transforms for burst denoising.
Liu et al. \cite{liu2021Swin} perform self-attention with shifted windows to boost the performance of vision transformer \cite{dosovitskiy2021an}. Recently, a series of MLP-based architectures \cite{msra_shift,AShift-MLP,S2-MLP} couple the spatial shifts with multi-layer perceptron to achieve competitive performances in high-level visions tasks. Liang et al. \cite{liang2022vrt} propose a video restoration transformer (VRT), where one video is partitioned into 2-frame clips at each layer and shifted for every other layer to perform temporal self-attention. However, it has a large number of self-attention layers and is computational costly. We extend shift operations to derive a large receptive field with small kernel convolutions.
\begin{figure*}
    \centering
    \includegraphics[width=0.86\textwidth]{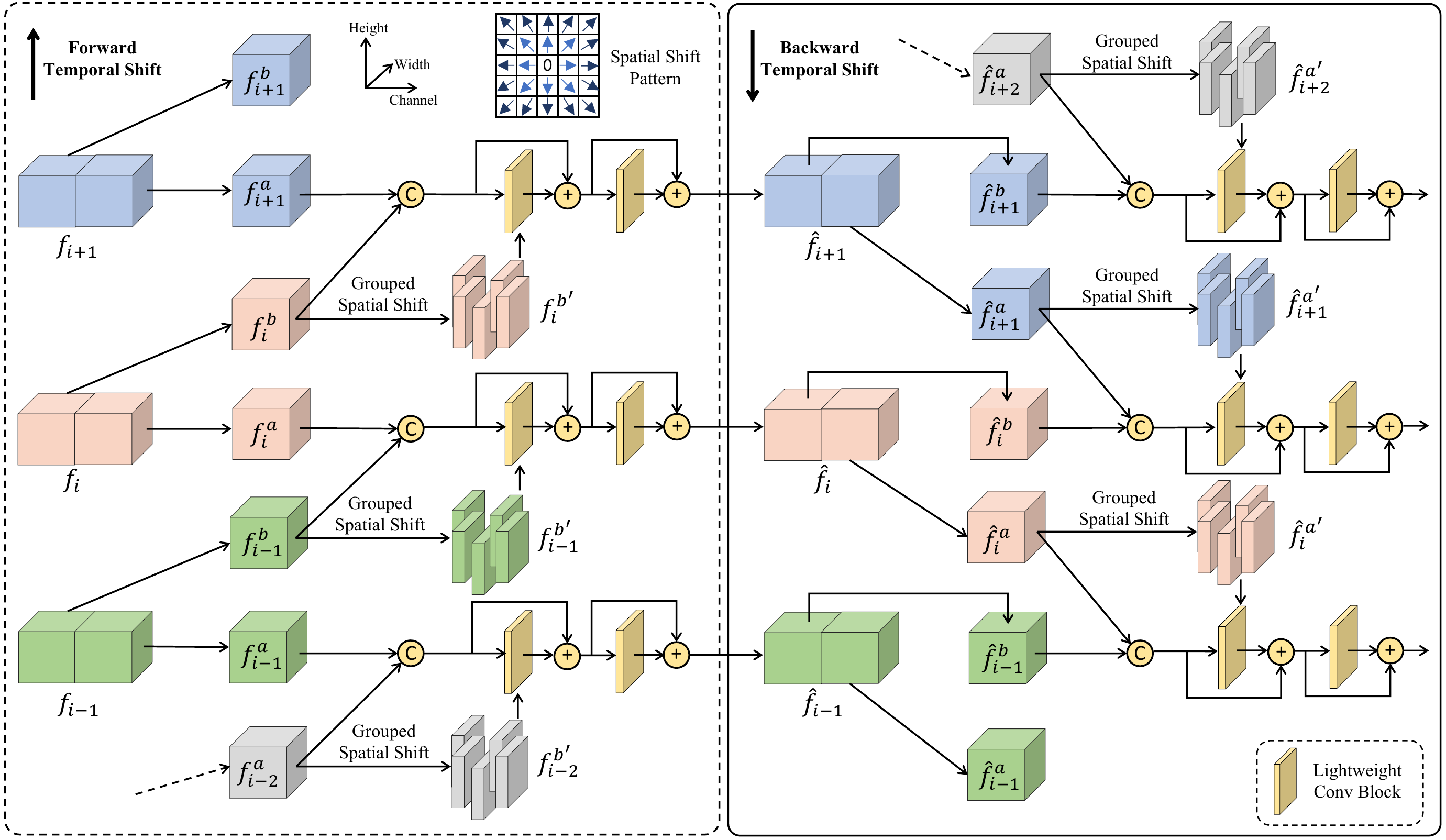}
    \vspace{-0.3cm}
    \caption{The operations of Grouped Spatial-temporal Shift (GSTS). We stack the forward temporal shift (FTS) blocks (\textit{Left}) and backward temporal shift (BTS) blocks (\textit{Right}) alternatively to achieve bi-directional propagation. Grouped spatial shift provides multiple candidate displacements within large spatial fields and establish temporal correspondences implicitly. 
    } 
    \label{fig:GSTS}
    \vspace{-0.4cm}
\end{figure*}
\section{Method}
Most previous methods in video restoration adopt complicated architectures, such as optical flow \cite{toflow}, deformable convolution \cite{dai2017deformable}, and self-attention layers \cite{liang2022vrt}. We propose a simple, yet effective grouped spatial-temporal shift block to establish temporal correspondences implicitly. 

\subsection{Overview of Group Shift-Net}
Given consecutive degraded frames $\{I_i\in\mathbb{R}^{h\times w \times c_{in}}\}_i^T$, where $T$ denotes the frame number, Group Shift-Net outputs the high-quality frames $\{O_i\in\mathbb{R}^{h\times w \times c_{out}}\}_i^T$. 
As shown in Fig~\ref{fig:overview}, our framework adopts a three-stage design: 1) feature extraction, 2) multi-frame feature fusion with grouped spatial-temporal shift, and 3) final restoration.

\noindent\textbf{Feature extraction.} Each frame $I_i$ usually suffers from different types of degradation (e.g. noise or blur),
which affects temporal correspondences modeling. Inspired by \cite{chan2022investigating}, 2D U-Net-like structures \cite{U-Net} are adopted to mitigate negative impacts of degradation and extract frame-wise features.

\noindent\textbf{Multi-frame feature fusion.}
At this stage, we propose a grouped spatial-temporal shift block to shift different features groups of neighboring frames to the reference frame to establish the temporal correspondences implicitly. 
The key-frame feature $f_i\in\mathbb{R}^{h\times w \times c}$ is fully aggregated with those of the neighboring frames to obtain the corresponding aggregated feature $A_i \in \mathbb{R}^{h\times w \times c}$.
Spatial-temporal shifts of different directions and distances are adopted to provide multiple candidate displacements for matching the frames.
By stacking multiple grouped spatial-temporal shift blocks, our framework can achieve long-term aggregation.

\noindent\textbf{Final restoration.} At last, U-Net-like structures take the low-quality input frames $\{I_i\}_i^T$ and corresponding aggregated features $\{A_i\}_i^T$ as input and produces each frame's final result $O_i$. 
The loss function $L$ is formulated as 
\begin{equation}
    L = \frac{1}{T} \sum_{i=1}^{T} ||H_i - O_i||_1.
\end{equation}

\subsection{Frame-wise Processing}
For feature extraction of stage 1 and final restoration of stage 3, 
we stack $N$ 2D slim U-Nets consecutively to extract features and conduct restoration effectively.
Stacking multiple U-Nets \cite{hourglass2016} was explored before, which leads to a deeper network depth and a larger receptive field than a single U-Net \cite{Zamir2021MPRNet} with the same computational cost. 
At each U-Net, we utilize residual blocks \cite{Resnet} to extract features. Average pooling and bilinear upsampling is adopted to adjust feature resolutions. The output features of the previous U-Net are directly passed to the next U-Net as input. 
The number $N$ and channels of stacked U-Nets are adjusted to meet different requirements of computational cost.  

\subsection{Grouped Spatial-temporal Shift}
In multi-frame fusion, frame-wise feature $f_i$ is aggregated with neighboring features $\{f_{i-t}, \dots, f_{i+t}\}$ to obtain temporally fused features $F_i$.
We adopt a 2D U-Net structure \cite{U-Net} for multi-frame fusion and keep skip connections in the U-Net.
We replace several 2D convolution blocks by stacking multiple grouped spatial-temporal shift (GSTS) blocks to effectively establish temporal correspondences and conduct multi-frame fusion. The GSTS blocks are not applied at the finest scale to save the computational cost.
A GSTS block consists of three components: 1) a temporal shift, 2) a spatial shift, and 3) a lightweight fusion layer, organized in the way shown in Figure~\ref{fig:GSTS}. 

\begin{table*}[t!]
  \small
  \centering
  \setlength{\tabcolsep}{5pt}
  \begin{tabular}{l|cccccccccccc}
    \toprule
    Method & EDVR & Su et al. & STFAN & TSP & MPRNet & MSDI & NAFNet & RNN-MBP & VRT & Ours-s & Ours & Ours+ \\
    \midrule
    PSNR  & 26.83 & 27.31 & 28.59 & 31.67 & 32.66 & 33.28 & 33.69 & 33.32 & 34.81 & 35.22 & 35.49 & \textbf{35.88} \\
    SSIM & 0.843 & 0.826 & 0.861 & 0.928 & 0.959 & 0.964 & 0.967 & 0.963 & 0.972 & 0.975 & 0.976 & \textbf{0.979} \\
    Params (M) & 20.6 & 15.3 & 5.37 & 16.17 & 20.1 & 241.3 & 67.8 & 16.4 & 18.3 & 4.1 & 10.5 & 12.3  \\
    FLOPS (G) & 194.2 & 38.7 & 35.4 & 357.9 & 760.1 & 336.4 & 63.3 & 496.0 & 721.3 & 47.1 & 146.5 & 151.3 \\
    \bottomrule
  \end{tabular}
  \vspace{-0.3cm}
  \caption{Quantitative comparison on GoPro \cite{deblur-multi-scale} test set.}
  \label{deblurring_flops_gopro}
  \vspace{-0.2cm}
\end{table*}
\setlength{\tabcolsep}{1pt}
\begin{figure*}[!t]
    \small
    \begin{center}
    \begin{tabular}{@{} c c @{}}
    \begin{tabular}{@{} c @{}}
               \includegraphics[width=.313\linewidth]{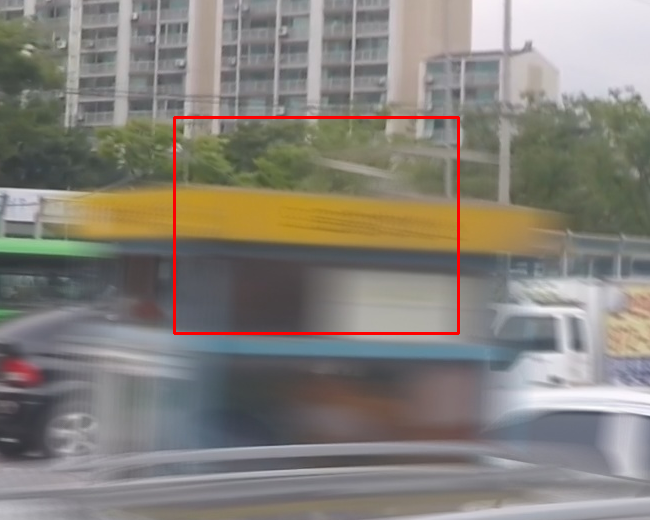} \\
               Video 410, Frame \textit{196} \\
               \includegraphics[width=.313\linewidth]{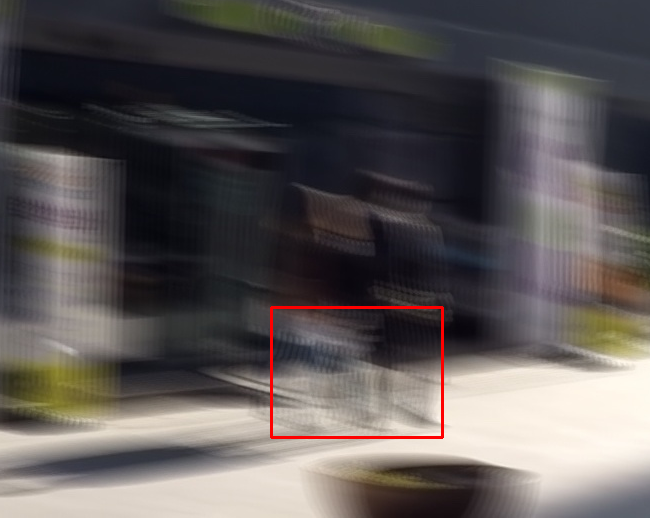} \\
               Video 881, Frame \textit{214} \\
    \end{tabular} & 
    \begin{tabular}{@{} c c c c @{}}
     \includegraphics[width=.145\linewidth]{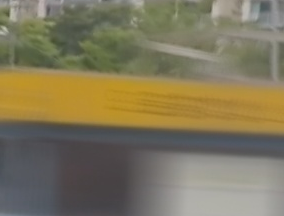} &
        \includegraphics[width=.145\linewidth]{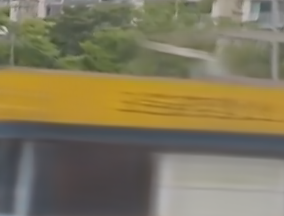} &
        \includegraphics[width=.145\linewidth]{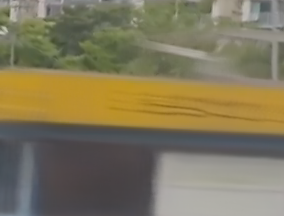} &
        \includegraphics[width=.145\linewidth]{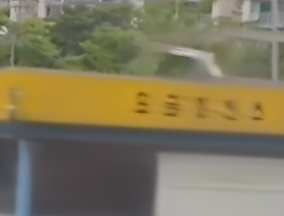} \\
        Input & TSP & MPRNet & NAFNet \\
        \includegraphics[width=.145\linewidth]{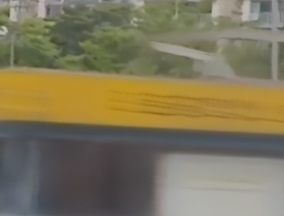} &
        \includegraphics[width=.145\linewidth]{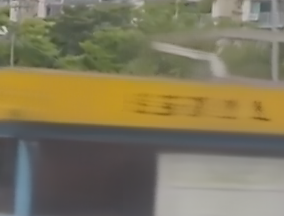} &
        \includegraphics[width=.145\linewidth]{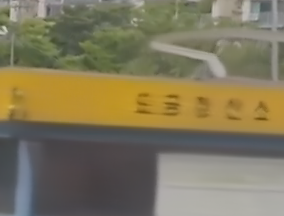} &
        \includegraphics[width=.145\linewidth]{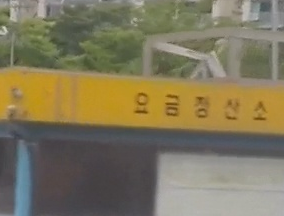} \\
        RNN-MBP & VRT & Ours & Ground Truth \\
        \includegraphics[width=.145\linewidth]{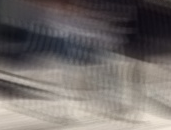} &
        \includegraphics[width=.145\linewidth]{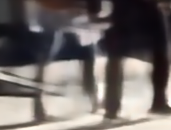} &
        \includegraphics[width=.145\linewidth]{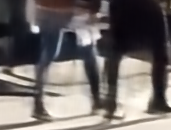} &
        \includegraphics[width=.145\linewidth]{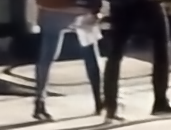} \\
        Input & TSP& MPRNet & NAFNet  \\
        \includegraphics[width=.145\linewidth]{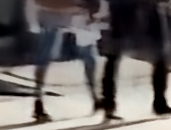} &
        \includegraphics[width=.145\linewidth]{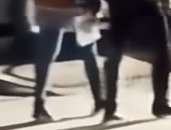} &
        \includegraphics[width=.145\linewidth]{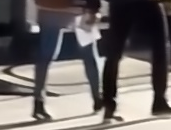} &
        \includegraphics[width=.145\linewidth]{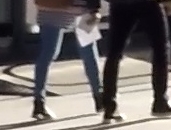} \\
        RNN-MBP & VRT & Ours & Ground Truth \\
    \end{tabular}
    \end{tabular}
    \end{center}
    \vspace{-0.7cm}
    \caption{Video deblurring on GoPro \cite{deblur-multi-scale} test set. Our method recovers more details than other methods.
    }
    \label{fig:vis_d}
    \vspace{-0.4cm}
\end{figure*}

\noindent\textbf{Grouped temporal shift.} It is observed in our experiment (Table~\ref{ab_shift}) that, 
handling three frames simultaneously \cite{lin2019tsm} would increase the difficulty of multi-frame fusion. To avoid it, our temporal shift processes only two adjacent frames. 
Grouped temporal shift blocks are either a forward temporal shift (FTS) block fusing $\{f_{i-1}, f_{i}\}$ (Figure~\ref{fig:GSTS} \textit{Left}) or a backward temporal shift (BTS) block fusing $\{f_{i+1}, f_{i}\}$ (Figure~\ref{fig:GSTS} \textit{Right}).
To achieve bi-directional aggregation, we stack FTS blocks and BTS blocks alternatively.% 

In a temporal shift, multi-frame features $f_i \in \mathbb{R}^{h\times w\times c}$ are split (i.e. grouped) equally along the channel dimension to obtain two feature groups: $f_i^a$ and $f_i^b$, where $f_i^{a},f_i^{b} \in \mathbb{R}^{h\times w\times \frac{c}{2}}$.
In the forward shift, $f_i^{a}$ is not shifted and is aggregated with the forward-shifted feature $f_{i-1}^{b}$ from time $i-1$. 
In the backward shift, $f_i^{a}$ is backward-shifted to be aggregated with $f_{i-1}^{b}$ for restoring $I_{i-1}$.
In other words, both FTS and BTS blocks keep half of the feature channels (one feature group) for characterizing visual appearance at current time $i$ and shift the other half of channels (the other feature group) for propagating information for inter-frame aggregation.
For simplicity, in the following paragraphs, we explain the details of the FTS block (i.e. how $f_i^{a}$ is aggregated with $f_{i-1}^{b}$), and the BTS block is similarly defined.

\noindent\textbf{Grouped spatial shift.} Concatenating $f_{i}^{a}$ and $f_{i-1}^{b}$ for restoring frame $i$ does not account for the spatial misalignment between two frames $i$ and $i$-1. 
Therefore, we perform additional spatial shift on the propagated feature group $f_{i-1}^{b} \in \mathbb{R}^{h\times w \times \frac{c}{2}}$ to achieve a large spatial range for spatial misalignment. Specifically, we first equally split (i.e. group) $f_{i-1}^{b}$ along the channel dimension to obtain $M$ feature slices $f_{i-1,m}^{b} \in \mathbb{R}^{h\times w \times \frac{c}{2M}}$, where $m=1,\dots,M$ is the slice index. For each feature slice $f_{i-1,m}^{b}$, we spatially shift it by ($\Delta x_m, \Delta y_m$) pixels in the $x$ and $y$ directions to obtain the shifted feature slice $f_{i-1,m}^{b'}$:
\begin{equation}
    f_{i-1,m}^{b'} = \mathrm{Shift}(f_{i-1,m}^{b},\Delta x_m,\Delta y_m). %~\mathrm{for}~ m=1,\dots,M
    \label{eq:shift_operation}
\end{equation}
$|\Delta x_m| =k_x*(s-1)+1, |\Delta y_m| =k_y*(s-1)+1$, where $k_x, k_y$ are integers and $s$ is defined as the base length of spatial shift.
When the spatial shift causes void pixels in the border, we set them to zero. 
For a $\Delta x_m$ pixels shift, the corresponding feature group is shifted spatially by $\Delta x_m$-1 pixels, followed by a depth-wise $3\times3$ convolution, which handles objects across two shifts and achieve smooth translation between two adjacent shifted feature slices.
Then we concatenate all feature groups $f_{i-1,m}^{b'}$ along the channel dimension to obtain the spatially shifted feature $f_{i-1}^{b'}$:
\begin{equation}
    f_{i-1}^{b'} = \mathrm{Concat}(f_{i-1,1}^{b'},\dots,f_{i-1,M}^{b'}).
\end{equation}
For example, when $M=9$ and $\Delta x_m, \Delta y_m \in \{-1,0,1\}$, the spatial shift operation creates $9$ feature slices and shifts the different slices by the 9 directions. In our implementation, we set $M=25$ and $\Delta x_m, \Delta y_m \in \{-9,-5,0,5,9\}$ to enlarge the alignment and fusion's receptive fields, so as to handle large displacements across frames.

\noindent\textbf{Fusion layer.} We utilize a fusion layer $F$ to aggregate multi-frame features $f_i^{a}, f_{i-1}^{b}, f_{i-1}^{b'}$.
The fusion layer $F$ contains two lightweight convolution blocks and each block adopts the combination between NAFNet \cite{chen2022simple} and Super Kernels \cite{sun2022skflow}, utilizing point-wise convolutions, depth-wise convolutions and gated layers to avoid heavy computation. 
The output fused feature $\hat{f_i}$ of frame $i$ is calculated as
\begin{equation}
    \hat{f_i} = \mathrm{Concat}(f_i^a, f_{i-1}^b) + \mathrm{F}(f_i^{a},f_{i-1}^{b}, f_{i-1}^{b'}).
    \label{eq:fusion_layer}
\end{equation}
The output feature $\hat{f_i}$ is fed to the next GSTS block.
To effectively merge shifted features, the kernel size of convolutions is set to be equal to the base shift length $s$.
\begin{figure}[b]
\vspace{-0.5cm}
\scriptsize
\centering
\setlength{\tabcolsep}{1pt}
\begin{tabular}{@{} c c c c @{}}
        \includegraphics[width=.2\linewidth]{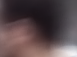} &
        \includegraphics[width=.2\linewidth]{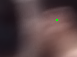} &
        \includegraphics[width=.2\linewidth]{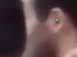} & 
        \includegraphics[width=.2\linewidth]{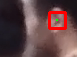} \\
        Input $I_{i-1}$ & Input $I_{i}$ & Output $O_i$ & GT $H_i$  \\
        \includegraphics[width=.2\linewidth]{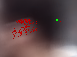} &
        \includegraphics[width=.2\linewidth]{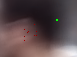} &
        \includegraphics[width=.2\linewidth]{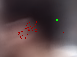} & 
        \includegraphics[width=.2\linewidth]{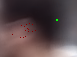} \\
        \textbf{LAM of shift $\rightarrow$} & LAM of shift $\leftarrow$ & LAM of shift $\uparrow$ & 
    LAM of shift $\downarrow$  \\
    \end{tabular}
    \vspace{-0.3cm}
    \caption{Local attribute visualization \cite{gu2021interpreting} of four shift directions. The saturation of red dots represent contribution weights of different areas in restoration of the marked local patch. 
    }
    \label{fig:vis_lam}
    \vspace{-0.5cm}
\end{figure}

\begin{table*}[t]
  \small
  \centering
  \setlength{\tabcolsep}{4.5pt}
  \begin{tabular}{l|ccccccccccccc}
    \toprule
    Method & EDVR & Su et al. & STFAN & TSP & PVDNet & ARVo & STDAN & ERDN & RNN-MBP & VRT & Ours-s & Ours & Ours+  \\
    \midrule
    PSNR  & 28.51 & 30.01 & 31.15 & 32.13 & 32.31 & 32.80 & 33.05 & 33.31 & 33.32 & 34.27 & 34.18 & 34.58 & \textbf{34.69}  \\
    SSIM & 0.864 & 0.887 & 0.905 & 0.927 & 0.926 & 0.935 & 0.937 & 0.940 & 0.963 & 0.965 & 0.965 & 0.968 & \textbf{0.969} \\
    \bottomrule
  \end{tabular}
  \vspace{-0.3cm}
  \caption{Quantitative comparison on DVD \cite{su2017deblurring} test set. }
  \label{deblurring_flops_dvd}
  \vspace{-0.2cm}
\end{table*}
\setlength{\tabcolsep}{1pt}
\begin{figure*}[!t]
    \scriptsize
    \begin{center}
    \begin{tabular}{@{} c c @{}}
    \begin{tabular}{@{} c @{}}
               \includegraphics[width=.234\linewidth]{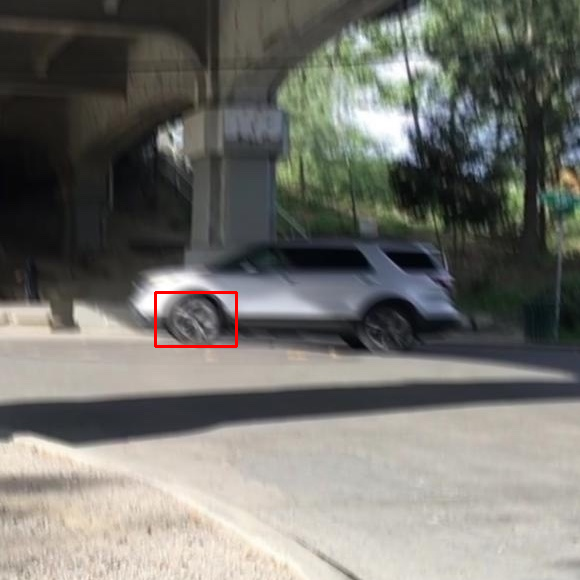} \\
               Video IMG\_0033, Frame \textit{30} \\
               \includegraphics[width=.234\linewidth]{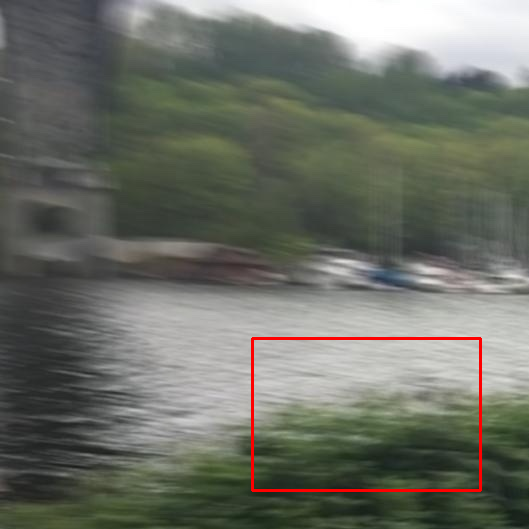} \\
               Video 720p\_240fps\_2, Frame \textit{70} \\
    \end{tabular} & 
    \begin{tabular}{@{} c c c c @{}}
    
        \includegraphics[width=.16\linewidth]{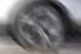} &
        \includegraphics[width=.16\linewidth]{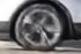} &
        \includegraphics[width=.16\linewidth]{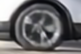} &
        \includegraphics[width=.16\linewidth]{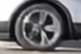} \\
        Input & PVDNet & TSP & STDAN\\
        \includegraphics[width=.16\linewidth]{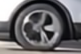} &
        \includegraphics[width=.16\linewidth]{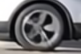} &
        \includegraphics[width=.16\linewidth]{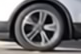} &
        \includegraphics[width=.16\linewidth]{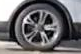}\\
        ERDN & VRT & Ours & Ground Truth \\
        \includegraphics[width=.16\linewidth]{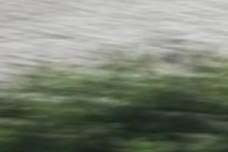} &
        \includegraphics[width=.16\linewidth]{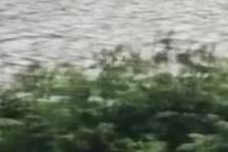} &
        \includegraphics[width=.16\linewidth]{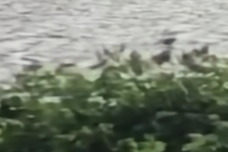} &
        \includegraphics[width=.16\linewidth]{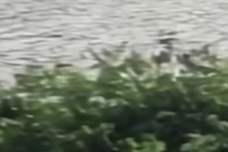}\\
        Input & PVDNet & TSP & STDAN \\
        \includegraphics[width=.16\linewidth]{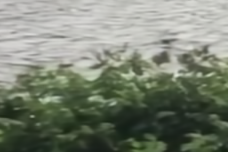} & 
        \includegraphics[width=.16\linewidth]{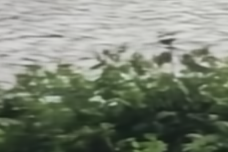} &
        \includegraphics[width=.16\linewidth]{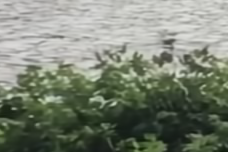} &
        \includegraphics[width=.16\linewidth]{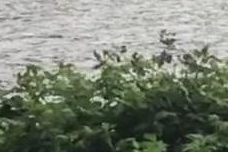} \\
        ERDN & VRT & Ours & Ground Truth \\
    \end{tabular}
    \end{tabular}
    \end{center}
    \vspace{-0.7cm}
    \caption{Video deblurring results on DVD \cite{su2017deblurring}. Our method performs better at reconstructing details of leaves and the moving tire.
    }
    \label{fig:vis_dvd}
    \vspace{-0.5cm}
\end{figure*}

\subsection{How Grouped Spatial Shift Help Restoration?}
We provide visualization and analysis to explore how different shifted features groups help video restoration. We input two neighboring frames into Group Shift-Net.
To analyze the feature map of grouped spatial shifts, we sample a $16\times16$ grid area from the resultant feature map.
Local attribution map (LAM) \cite{gu2021interpreting} is performed to analyze contribution weights of all shifted features in restoring the $16\times16$ grid. The contribution weights are visualized as the red dots in Figure~\ref{fig:vis_lam}. When the color of dots is more saturated, the local area is more important in restoration. 
It is shown that the shifted features are more important in restoring $O_i$, when a shift direction is similar to the motion between $I_{i-1}$ and $I_i$. Moreover, our method could obtain expansive effective receptive fields for temporal correspondence establishment.

\section{Experiments}

We conduct experiments and ablation study on two tasks: video deblurring and video denoising.

\noindent\textbf{Datasets.} For video deblurring, we train and evaluate our method on GOPRO~\cite{deblur-multi-scale} and DVD~\cite{su2017deblurring} datasets. %which are synthesized by averaging consecutive images from high-speed camera. 
GOPRO~\cite{deblur-multi-scale} dataset contains 2,103 and 1,111 frames as training and test sets, respectively. DVD~\cite{su2017deblurring} includes 5,708 frames for training and 1,000 frames for testing. For video denoising, we follow Huang et al. \cite{Compression2022} to train our model with noise level $\sigma\in\mathcal{U}\left[0,50\right]$ on DAVIS~\cite{khoreva2018video} dataset and test it on DAVIS \cite{khoreva2018video} of different noise levels. 
\setlength{\tabcolsep}{3.3pt}
\begin{table*}
\footnotesize
\small
  \centering
  \begin{tabular}{cc|cccccccccccc}
    \toprule
Dataset  & $\sigma$ & VLNB & DVDNet & FastDVD & EMVD-L & PaCNet & Huang et al. & FloRNN & Tempformer & VRT & Ours-s & Ours & Ours+ \\ 
\midrule
\multirow{5}{*}{DAVIS} & 10 & 38.85 & 38.13 & 38.71 & 38.57 & 39.97  & 39.67 & 40.16 & 40.17 & 40.82 & 40.55 & 40.75 & \textbf{40.91} \\
 & 20 & 35.68 & 35.70 & 35.77 & 35.39 & 36.82 & 36.33 & 37.52 & 37.36 & 38.15 & 37.84 & 38.19 & \textbf{38.34} \\
 & 30 & 33.73 & 34.08 & 34.04 & 33.89 & 34.79 & 34.62 & 35.89 & 35.66 & 36.52 & 36.25 & 36.62 & \textbf{36.83} \\
 & 40 & 32.32 & 32.86 & 32.82 & 32.40 & 33.34 & 33.40 & 34.66 & 34.42 & 35.32 & 35.11 & 35.47 & \textbf{35.71} \\
 & 50 & 31.13 & 31.85 & 31.86 & 31.47 & 32.20 & 32.41 & 33.67 & 33.44 & 34.36 & 34.20 & 34.53 & \textbf{34.82} \\ \midrule
 \multicolumn{2}{c|}{Params (M)} & - & - & 2.5 & 9.6 & 2.87 & 13.95 & 11.8 & - & 18.3 & 3.7 & 10.8 & 12.7 \\
 \multicolumn{2}{c|}{FLOPS (G)} & - & - & 41.8 & 69.5 & - & 48.5 & 189.7 & - & 721.3  & 47.2 & 146.8 & 154.3 \\
    \bottomrule
  \end{tabular}
  \vspace{-0.35cm}
  \caption{Quantitative comparison on DAVIS~\cite{khoreva2018video} test set.}
  \label{set8_results}
  \vspace{-0.2cm}
\end{table*}
\setlength{\tabcolsep}{1pt}
\begin{figure*}[t]
    \small
    \begin{center}
    \begin{tabular}{@{} c c @{}}
    \begin{tabular}{@{} c @{}}
               \includegraphics[width=.321\linewidth]{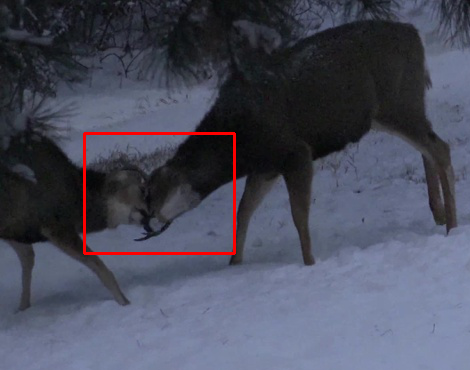} \\
               Video deer, Frame \textit{10} \\
               \includegraphics[width=.321\linewidth]{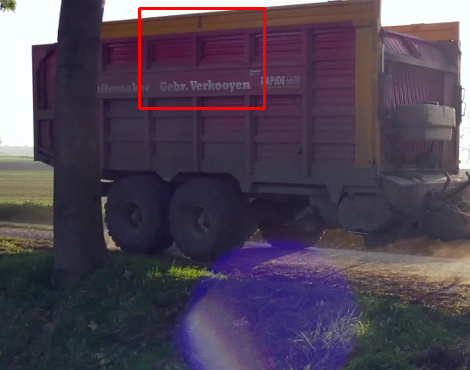} \\
               Video tractor, Frame \textit{5} \\
    \end{tabular} & 
    \begin{tabular}{@{} c c c c @{}}
        \includegraphics[width=.14\linewidth]{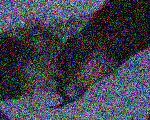} &
        \includegraphics[width=.14\linewidth]{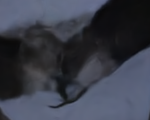} &
        \includegraphics[width=.14\linewidth]{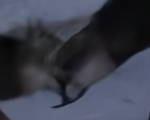} &
        \includegraphics[width=.14\linewidth]{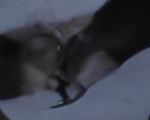} \\
        Input & DVDNet & FastDVD & FloRNN \\
        \includegraphics[width=.14\linewidth]{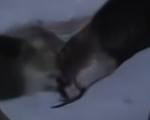} &
        \includegraphics[width=.14\linewidth]{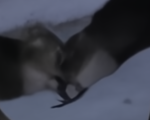} &
        \includegraphics[width=.14\linewidth]{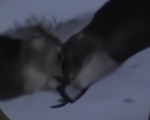} &
        \includegraphics[width=.14\linewidth]{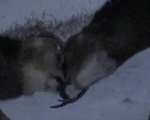} \\
        PaCNet & VRT & Ours & Ground Truth \\
        \includegraphics[width=.14\linewidth]{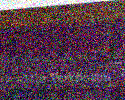} &
        \includegraphics[width=.14\linewidth]{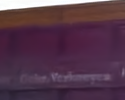} &
        \includegraphics[width=.14\linewidth]{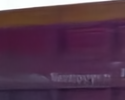} &
        \includegraphics[width=.14\linewidth]{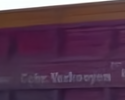}
        \\
        Input & DVDNet & FastDVD & FloRNN \\
        \includegraphics[width=.14\linewidth]{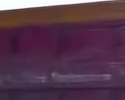} &
        \includegraphics[width=.14\linewidth]{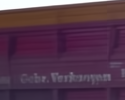} &
        \includegraphics[width=.14\linewidth]{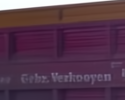} &
        \includegraphics[width=.14\linewidth]{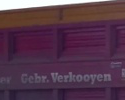} \\
        PaCNet & VRT & Ours & Ground Truth \\
    \end{tabular}
    \end{tabular}
    \end{center}
    \vspace{-0.7cm}
    \caption{Video denoising results on DAVIS~\cite{khoreva2018video} test set. Our method reconstructs more details of textures and texts.
    }
    \label{fig:vis_denoising}
    \vspace{-0.5cm}
\end{figure*}

\noindent\textbf{Model Scaling.} We provide small model (denoted as ``Ours-s''), base model (denoted as ``Ours'') to meet different computational requirements. For both models, We replace the convolution blocks by multiple GSTS blocks in Stage-2's UNet. We further observe that merely replacing convolution blocks in decoders of Stage-2's UNet (denoted as ``Ours+'') could boost the performances further.
The details of different models are in Appendix.  

\noindent\textbf{Implementation details.} 
Our network is end-to-end trained. 
The base shift length $s$ is set to $5$. The networks are trained with a batch size of 8 for 750 epochs. The re-parameterization technique \cite{ding2021repvgg} is adopted to optimize convolutions in GSTS. The patch size is set as $256\times256$. Horizontal and vertical flips are adopted for data augmentation. We use the Adam optimizer \cite{adam} and the learning rate is decreased from $4\times10^{-4}$ to $1\times10^{-7}$ according to the cosine annealing strategy \cite{cosine_annealing}. At inference of video deblurring, ``Ours-s'' processes 100 frames simultaneously. ``Ours'' and ``Ours+'' process only 50 frames due to the memory limit.

\subsection{Video Deblurring Results} 
\noindent\textbf{Quantitative comparison.} 
We compare our method with state-of-the-art deblurring methods including EDVR  \cite{wang2019edvr}, Su et al. \cite{su2017deblurring}, STFAN \cite{zhou2019stfan}, TSP \cite{Pan2020cdvdtsp}, MPRNet \cite{Zamir2021MPRNet}, MSDI-Net \cite{Li2022Learning}, NAFNet \cite{chen2022simple}, RNN-MBP \cite{RNN-MBP}, STDAN \cite{zhang2022spatio}, ERDN \cite{ERDN} and VRT \cite{liang2022vrt}.
As shown in Tables~\ref{deblurring_flops_gopro} and~\ref{deblurring_flops_dvd}, ``our+'' outperforms VRT~\cite{liang2022vrt}, the most competitive method, by 1.07 dB and 0.42 dB PSNR on GoPro and DVD, respectively, with only 21\% of its flops. For a more intuitive comparison, we provide the PSNR-Params-FLOPS plot in Figure~\ref{fig:vis_flops_psnr}. The two versions of our model occupy the top-left corner, showing the best performances with less computational cost. Notably, ``Ours-s'' surpasses STFAN~\cite{zhou2019stfan} by a significant 6.63 dB PSNR with the fewest parameters.

\noindent\textbf{Qualitative comparison.} 
Figure ~\ref{fig:vis_d} provides the visualization of two hard deblurring cases. As one can see from the full images, there exist severe blurry regions due to camera shaking and object movement. On the zoomed-in patchs, our model reconstructs much sharper letters, building structures and boundaries of moving legs.
\subsection{Video Denoising Results}
\noindent\textbf{Quantitative comparison.} We compare our method with SOTA video denoising methods VLNB \cite{VLNB}, DVDNet \cite{dvdnet}, FastDVD \cite{fastdvdnet}, EMVD-L \cite{EMVD}, PaCNet \cite{Vaksman2021PatchC}, Huang et al. \cite{Compression2022}, FloRNN \cite{li2022unidirectional}, Tempformer \cite{tempformer} and VRT \cite{liang2022vrt}.
It is shown in Table~\ref{set8_results} that we achieve best performances in 5 noise levels on with less computational cost. Moreover, our small model performs better than previous lightweight models, such as FastDVD \cite{fastdvdnet}, EMVD-L \cite{EMVD}.

\noindent\textbf{Qualitative comparison.} 
Figure ~\ref{fig:vis_denoising} visualizes the denoising results of DAVIS~\cite{khoreva2018video} . Note the zoomed-in regions in the red boxes. Other models generate over-smooth results, while our model reconstructs more details in grass and texts.

\subsection{Ablation Study}
We demonstrate the effectiveness of each key component in Group Shift-Net. All compared methods are trained and evaluated with the same training settings of our base model.

\noindent\textbf{Spatial Temporal shift.} 
 We evaluate the impact of \textit{grouped spatial shift} and \textit{alternative temporal shift} in Table~\ref{ab_shift}. 
 At first, We remove \textit{grouped spatial shift} and merely apply alternative temporal shift . 
 The kernel size of convolution in the fusion layers is set to be $3\times3$, which is widely used previously \cite{lin2019tsm,tsmdenoising}. It suffers a drop of 0.35 dB PSNR. 
 Then we replace alternative temporal shift by bi-directional shift, where the fusion layer would aggregate $\frac{3}{4}$ channels of feature $f_i$, $\frac{1}{8}$ channels of feature $f_{i-1}$, and $\frac{1}{8}$ channels of feature $f_{i+1}$. This operation causes a decrease of 0.33 dB PSNR. 
 The ablation illustrates the importance of \textit{grouped spatial shift} and \textit{alternative temporal shifts}.

\setlength{\tabcolsep}{4pt}
\begin{table}
 \scriptsize
  \centering
  \begin{tabular}{cc|c}
  \hline 
             \makecell[c]{Alternative \\ Temporal Shift} & \makecell[c]{Spatial \\ Shift}  & PSNR \\ \hline
             \xmark & \xmark & 34.81  \\
             \cmark & \xmark & 35.14  \\ \hline
             \cmark & \cmark & \textbf{35.49}  \\
            \hline
          \end{tabular}
  \vspace{-0.3cm}
  \caption{Ablation of grouped spatial-temporal shift.}
  \label{ab_shift}
  % \vspace{-0.1cm}
  \setlength{\tabcolsep}{2pt}
  \begin{tabular}{l|ccccc}
                \hline
                Receptive Field & $5\times5$ & $9\times9$ & $17\times17$ & $25\times25$ & $33 \times 33$ \\ \hline
                Ours w/o spatial & 35.14 & 35.20 & 35.18 & 35.16 & 35.17 \\ 
                \hline
  \end{tabular}
  \begin{tabular}{l|cccc}
                \hline
                Receptive Field & $13\times13$ & $23\times23$ & $33 \times 33$ \\ \hline
                Ours & 35.39 & \textbf{35.49} & 35.48 \\ 
                \hline
  \end{tabular}
  \vspace{-0.3cm}
  \caption{Receptive field and spatial shift in a fusion layer.}
  \label{ab_largeKernel}
  % \vspace{0.1cm}
  \setlength{\tabcolsep}{1pt}
  \begin{tabular}{c|cccc}
                \hline
                $\Delta x_m, \Delta y_m$ & $\{0\}$ & $\{0,\pm1\}$ & $\{0,\pm2,\pm3\}$ & $\{0,\pm3,\pm5\}$ \\ \hline
                PSNR & 35.20 & 35.29 & 35.37 & 35.35  \\ \hline
                $\Delta x_m, \Delta y_m$ & $\{0,\pm4,\pm7\}$ & $\{0,\pm5,\pm9\}$ & $\{0,\pm6,\pm11\}$ & $\{0,\pm7,\pm13\}$ \\ \hline
                PSNR & 35.44 & \textbf{35.49} & 35.46 & 35.40  \\
                \hline
          \end{tabular}
  % \vspace{0.1cm}
  \setlength{\tabcolsep}{2pt}
  \begin{tabular}{c|cccc}
                \hline
                $\Delta x_m, \Delta y_m$ & $\{0\}$ & $\{0,\pm5\}$ & $\{0,\pm5,\pm9\}$ & $\{0,\pm5,\pm9,\pm13\}$ \\ \hline
                PSNR & 35.14 & 35.34 & \textbf{35.49} & 35.47 \\ 
                \hline
          \end{tabular}
          \vspace{-0.3cm}
          \caption{Ablation of $(\Delta x_m, \Delta y_m)$ in grouped spatial shift.}
          \label{ab_spatial_shift2}
  \vspace{-0.5cm}
\end{table}

\noindent\textbf{Receptive field in fusion layers.} We change the base shift length $s$ to be $3,5,7$. The corresponding receptive fields of a fusion layer (depth-wise convolutions with kernel size $s$+1) are $13\times13,23\times23,33\times33$. 
We also remove spatial shift and enlarge the kernel sizes of convolutions to achieve similiar receptive field (denoted as ``Ours w/o spatial''). The kernel sizes of the depth-wise convolution are set to be 3, 5, 9, 13, 17 and the corresponding receptive fields are $5\times5, 9\times9,17\times17,25\times25,33\times33$, respectively. It is shown in Table~\ref{ab_largeKernel} that larger kernel convolutions cannot achieve better performances. It might be because extremely large kernels are stiill difficult to optimize. It is also observed that our method surpasses optimizing larger kernels by about 0.3 dB PSNR, which demonstrates the superiority of spatial shift. 
 
\noindent\textbf{Grouped spatial shift.}
We first set $M=25$ and set the kernel size in fusion layers to be 5. 
Then we change the base shift length $s$ from 1 to 7 and $\Delta x_m, \Delta y_m$ as shown in Table~\ref{ab_spatial_shift2}.
$\Delta x_m, \Delta y_m \in \{0\}$ means that only temporal shift is applied. 
It is observed that the model with $\Delta x_m, \Delta y_m \in \{0,\pm5,\pm9\}$ achieves the best performance. 
When the base shift length $s$ increases, the spatial shifts with larger receptive fields achieve better performance. The models suffer degraded performances when the shift length is larger than the kernel size. It is because convolutions would not filter shifted features seamlessly. Then we change the number $M$ of shifts and $\Delta x_m, \Delta y_m$ are changed with the number $M$. It is shown in Table~\ref{ab_spatial_shift2} that the models with $M=49$ and $M=25$ achieve the similar performances, which outperform the model with $M=9$ by about 0.15 dB PSNR. 

\noindent\textbf{Temporal consistency.} 
Following Tempformer \cite{tempformer}, we add noise with 12 different noise seeds on DAVIS to create a dataset of 12-frame sequences.
Mean absolute error between adjacent outputs is taken as the metric. Table~\ref{tab:temporal_consistent} shows that our method and VRT achieve similar consistency.

\noindent\textbf{Replacing shift blocks with optical flow, DCN and self-attention.} We first replace our fusion layer in shift blocks with DCN (denoted as ``GSTS+DCN'') and cross-frame (shifted window size=8) self-attention layers (denoted as ``GSTS+self-attn''). Table~\ref{tab:re3_q2} shows that our simple structure achieves better performance than DCN and self-attention. Then we replace shift blocks with optical flow (a pre-trained SPyNet as initialization), DCN layers and cross-frame self-attention layers (shifted window size = 8), separately. It is observed in Table~\ref{tab:re3_q2} that our method achieves better performance with less computational cost. 

\noindent\textbf{Shift length $s$.} We first evaluate shift length $s$ on different types of degradation, such as blur and noise. It is observed in Table~\ref{tab:shift_s} that the network with $s$=5 achieves the best performance on both video deblurring and denoising. We further apply bicubic upsampling and bilinear downsampling on DAVIS (480$\times$854) to obtain a downsampled DAVIS dataset (240$\times$427) and a upsampled DAVIS dataset (960$\times$1708). As shown in Table~\ref{tab:shift_s}, the network with $s$=5 achieves the best performance at all resolutions.

\begin{table}[t]
    \scriptsize
    \centering
    \begin{tabular}{c|ccc}
    % \toprule
    \hline 
        Method & $\sigma=10$ & $\sigma=30$ & $\sigma=50$ \\ \hline
        VRT & $1.5 \times 10^{-3}$ & $1.6 \times 10^{-3}$ & $2.0 \times 10^{-3}$ \\ \hline
        Ours & $1.7 \times 10^{-3}$ & $1.7 \times 10^{-3}$  & $1.9 \times 10^{-3}$ \\ 
        % \bottomrule
    \hline
    \end{tabular}
    \vspace{-0.3cm}
    \caption{Temporal consistency evaluation of video denoising.}
    \label{tab:temporal_consistent}
\setlength{\tabcolsep}{3pt}
    \scriptsize
    \centering
    \begin{tabular}{c|c|ccc|cc}
    % \toprule
    \hline
    \multirow{2}{*}{Method}  & Deblurring & \multicolumn{3}{c|}{DAVIS denoising} & \multirow{2}{*}{Params} & \multirow{2}{*}{FLOPs} \\ \cline{2-5}
    & GoPro & $\sigma$=10 & $\sigma$=30 & $\sigma$=50& &  \\ \hline
    GSTS + self-attn & 34.67 & 40.54 & 36.28 & 34.17 & 11.1 (M) & 168.7 (G) \\
        GSTS + DCN & 33.74 & 40.02 & 35.65 & 33.43 & 17.9 (M) & 210.3 (G)   \\ \hline
        Optical Flow & 34.14 & 40.07 & 35.68 & 33.55 & 14.2 (M) & 189.4 (G) \\
        self-attn & 34.52 & 40.58 & 36.31 & 34.17 & 10.6 (M) & 153.6 (G) \\
        DCN &  33.66 & 39.91 & 35.48 & 33.18 & 17.2 (M) & 203.8 (G)  \\  \hline
        Ours & \textbf{35.49} & \textbf{40.75} & \textbf{36.62} & \textbf{34.53} & 10.8 (M) & 146.8 (G) \\ 
        \hline % \bottomrule
    \end{tabular}
    \vspace{-0.3cm}
    \caption{Replacing shift blocks with several variants.}
    \label{tab:re3_q2}
    \centering
\setlength{\tabcolsep}{2pt}
    \begin{tabular}{c|c|ccc|ccc|ccc}
    % \toprule
    \hline 
        \multirow{2}{*}{$s$} & Deblurring & \multicolumn{3}{c|}{DAVIS (480$\times$854)} & \multicolumn{3}{c|}{DAVIS (240$\times$427)} & \multicolumn{3}{c}{DAVIS (960$\times$1708)} \\ \cline{2-11}
        & GoPro & $\sigma$=10 & $\sigma$=30 & $\sigma$=50 & $\sigma$=10 & $\sigma$=30 & $\sigma$=50 & $\sigma$=10 & $\sigma$=30 & $\sigma$=50  \\ \hline
        3 & 35.39 & 40.65 & 36.47 & 34.34 & 39.07 & 34.71 & 32.65  & \textbf{43.20} & 38.92 & 36.93  \\ \hline
        5 & \textbf{35.49} & \textbf{40.75} & \textbf{36.62} & \textbf{34.53}  & \textbf{39.13} & \textbf{34.82} & \textbf{32.77} & \textbf{43.20} & \textbf{39.05} & \textbf{37.11} 
        \\ \hline
        7& 35.48 & 40.63 & 36.45 & 34.32 & 39.05 & 34.69 & 32.62 & 43.19 & 38.94 & 36.98 \\ \hline
        9& 35.33 & 40.56 & 36.35 & 34.20 & 39.01 & 34.61 & 32.54 & 43.19 & 38.86 & 36.84 \\
       %  \bottomrule
       \hline
    \end{tabular}
    \vspace{-0.3cm}
    \caption{Shift length $s$ on different degradation and resolutions.}
    \label{tab:shift_s}
    \vspace{-0.6cm}
\end{table}

\section{Conclusion}
In this paper, we propose a simple and effective framework for video restoration that does not require complicated architectures like optical flow, deformable convolution, or self-attention. Instead, we introduce a simple spatial temporal shift block for implicit temporal correspondence modeling. Our method outperforms state-of-the-art methods with less computational cost on video deblurring and denoising tasks. We do not foresee any negative social impact resulting from this work.
\setcounter{section}{0}
\section*{Appendix}
\section{Optical Flow Analysis in Video Restoration}
Optical flow, the core component to model the motion information, has been widely used in video super-resolution \cite{chan2021basicvsr, chan2021basicvsr++}, video deblurring \cite{Son2021PVDNet, Pan2020cdvdtsp} and video denoising \cite{toflow}. However, Zhu et al. \cite{RNN-MBP} demonstrate that optical flow cannot estimate the alignment information well because of the significant influence of the motion blur. \cite{SeeMotion} also show that the optical flow is not accurate in noisy images.

We provide a quantitative analysis of optical flow in three video restoration tasks, including video super-resolution, video deblurring and video denoising.
We select BasicVSR++ \cite{chan2021basicvsr++} (denoted as ``BasicVSR++'') as the baseline model. To evaluate the importance of optical flow module, we remove the optical flow estimation from BasicVSR++ (denoted as ``BasicVSR++ w/o flow''). We increase the number of residual blocks \cite{Resnet} and offsets computing layers in DCN to maintain the same computational cost as BasicVSR++. 
Then two models are trained for 200,000 iterations on video super-resolution (REDS4 dataset \cite{reds4}), video deblurring (GoPro dataset \cite{deblur-multi-scale}) and video denoising (Set8 dataset \cite{dvdnet}), respectively. 
For a fair comparison of three tasks, we do not take the generalized verison \cite{chan2022generalization} of BasicVSR++ and the models for video deblurring and video denoising have the same parameters as BasicVSR++ \cite{chan2021basicvsr++} for video super-resolution. It is observed in Table~\ref{basicvsrpp_flow} that the optical flow module makes different influences on different tasks. The optical flow can boost the performance of super-resolution by 0.56 dB. However, optical flow \textbf{cannot improve video deblurring and denoising greatly} because optical flow is not that accurate in blurry and noisy images as shown in \cite{SeeMotion,RViDeNet,Son2021PVDNet,RNN-MBP}.

\begin{figure*}[!t]
    \scriptsize
    \begin{center}
    \begin{tabular}{@{} c c c @{}}
         \includegraphics[width=.32\linewidth]{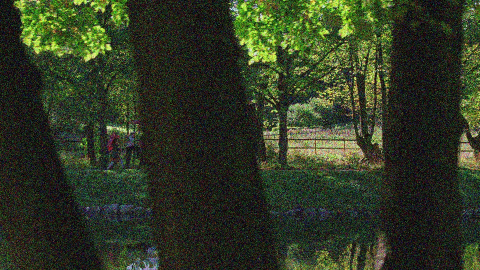} &
        \includegraphics[width=.32\linewidth]{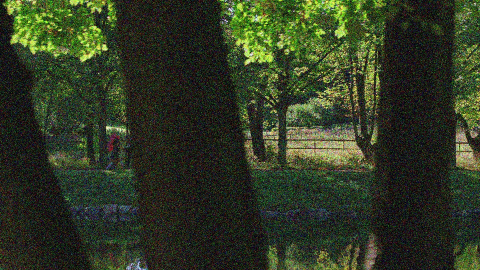} &
        \includegraphics[width=.32\linewidth]{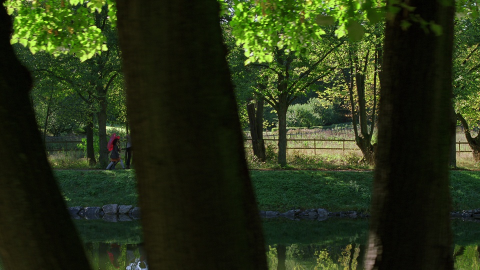} \\
        Noisy frame $I_{i-1}$  & Noisy frame $I_i$ & Clean frame $GT_i$  \\
        \includegraphics[width=.32\linewidth]{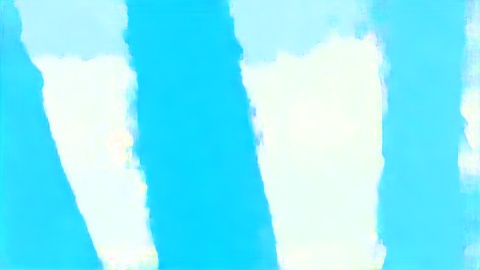} &
        \includegraphics[width=.32\linewidth]{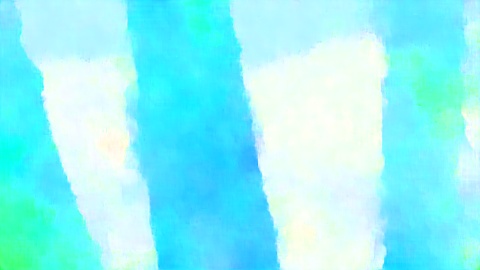} & 
        \includegraphics[width=.32\linewidth]{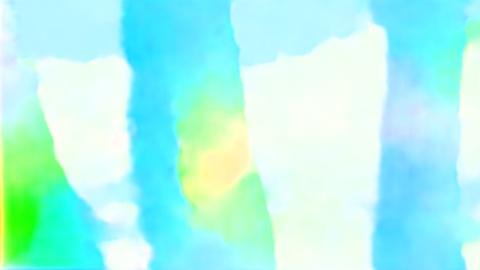} \\
        Optical flow from clean pairs & Optical flow from noisy pairs & Task-oriented flow \cite{chan2022generalization} from noisy pairs   \\
        \includegraphics[width=.32\linewidth]{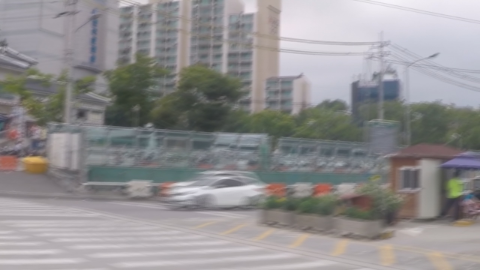} &
        \includegraphics[width=.32\linewidth]{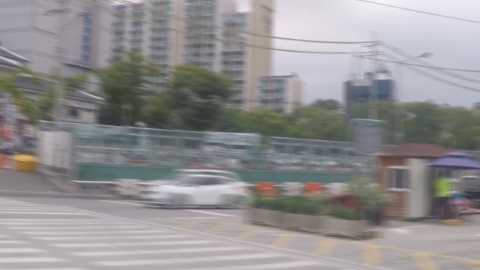} &
        \includegraphics[width=.32\linewidth]{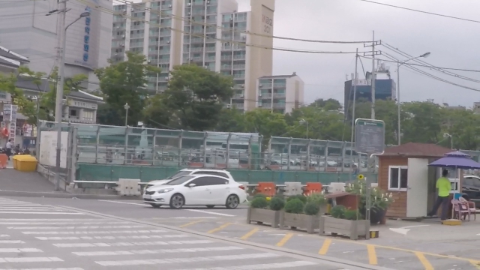} \\
        Blurry frame $I_{i-1}$  & Blurry frame $I_i$ & Sharp frame $GT_i$  \\
        \includegraphics[width=.32\linewidth]{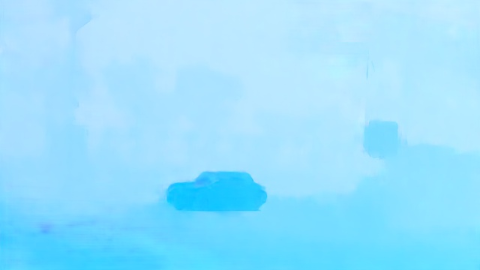} &
        \includegraphics[width=.32\linewidth]{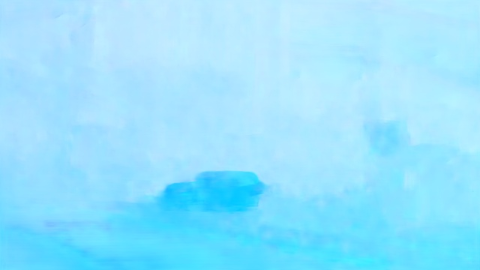} & 
        \includegraphics[width=.32\linewidth]{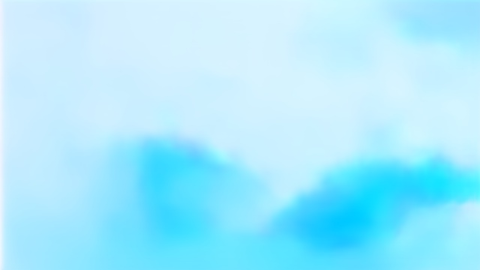} \\
        Optical flow from sharp pairs & Optical flow from blurry pairs & Task-oriented flow \cite{chan2022generalization} from blurry pairs   \\
    \end{tabular}
    \end{center}
    \vspace{-0.7cm}
    \caption{Optical flow visualization on GoPro testset \cite{deblur-multi-scale} and Set8 testset \cite{dvdnet}. The optial flow estimation of blurry images and noisy images is not accurate due to the negative influence of blur and noise. Task-oriented flow is usually smooth and inaccurate.
    }
    \label{fig:vis_flow}
    \vspace{-0.3cm}
\end{figure*}

We also provide a visualization of optical flow in Figure~\ref{fig:vis_flow}. Given degraded input frames $I_{i-1},I_{i}$ and ground truth frames $GT_{i-1}, GT_{i}$, we utilize a pre-trained optical flow model \cite{Spynet} to estimate the optical flow of degraded pairs $I_{i} \rightarrow I_{i-1}$ and ground truth pairs $GT_{i} \rightarrow GT_{i-1}$. We also utilize the optical flow network trained in the generalized version \cite{chan2022generalization} of BasicVSR++ to visualize the task-oriented flow. It is shown in Figure~\ref{fig:vis_flow} that the optical flow estimation is not accurate in noisy frames or blurry frames. Even with training on GoPro dataset \cite{deblur-multi-scale} or DAVIS dataset~\cite{khoreva2018video}, the task-oriented flow could not produce more accurate optical flow.

The different influences of optical flow estimation illustrate that the optical flow could help improve video super-resolution but make small contribution to video deblurring and denoising. 
Since it is difficult for optical flow to model motion information directly in video deblurring and video denoising, we design grouped spatial-temporal shift to achieve large receptive fields for \textbf{implicit temporal correspondence modeling when optical flow is inaccurate}. The network is not designed for video super-resolution, which optical flow estimation could greatly help. Our network does not utilize optical flow and may not perform well on video super-resolution.

\section{More Results}
\begin{table}[t]
\footnotesize
\setlength{\tabcolsep}{2pt}
\centering
  \begin{tabular}{c|c|c|ccc|c}
    \toprule
     \multirow{2}{*}{Method}  & SR & Deblurring & \multicolumn{3}{c|}{Denoising} & \multirow{2}{*}{Params}  \\ \cline{2-6}
     & REDS4 & GoPro & $\sigma$=10 & $\sigma$=30 & $\sigma$=50 & \\ \hline
    BasicVSR++ & \textbf{32.01}  & 33.22 & 36.19 & 31.75 & 29.56 & 7.3M  \\ \hline
    BasicVSR++ w/o flow & 31.45 & 33.25 & 36.10 & 31.62 & 29.49 & 6.6M \\
    \bottomrule
  \end{tabular}
  \vspace{-0.3cm}
  \caption{Analysis of optical flow on different tasks. Optical flow can improve video super-resolution greatly (+0.56 dB PSNR), but cannot in video deblurring and denoising.}
  \label{basicvsrpp_flow}
  \vspace{-0.5cm}
\end{table}
\noindent\textbf{Qualitative visualization.} We provide four videos (Deblurring1.mp4, Deblurring2.mp4, Denoising1.mp4, Denoising2.mp4) in the project pages. Deblurring1.mp4 and Denoising1.mp4 provide the full-frame visualization of our restored video. It is shown that ours videos do not produce flicker cases and are temporal consist and stable. Deblurring2.mp4 and Denoising2.mp4 are provided to compare VRT \cite{liang2022vrt} and our method clearly. It is shown that our method can restore more textures and details than VRT in both video deblurring and video denoising. 

\noindent\textbf{Set8.} We also evaluate our model of video denoising on set8 \cite{dvdnet}. The results are shown in Table~\ref{set8_result}. Our method achieves the approximate performances to a concurrent method: RVRT~\cite{liang2022rvrt}.

\setlength{\tabcolsep}{0.5pt}
\begin{table*}
\footnotesize
\small
\setlength{\tabcolsep}{3.3pt}
  \centering
  \begin{tabular}{cc|cccccccccccc}
    \toprule
Dataset  & $\sigma$ & VLNB & DVDNet & FastDVD & EMVD-L & PaCNet & Huang et al. & FloRNN & Tempformer & VRT & RVRT & Ours-s & Ours+ \\ 
\midrule
\multirow{5}{*}{Set8} & 10 & 37.26 & 36.08 & 36.44 & 36.56 & 37.06 & 37.17 & 37.57 & 37.15 & 37.88 & 37.53 & 37.43 & 37.48 \\
& 20 & 33.72 & 33.49 & 33.43 & 33.27 & 33.94 & 34.22 & 34.67 & 34.74 & 35.02 & 34.83 & 34.53 & 34.85 \\
& 30 & 31.74 & 31.79 & 31.68 & 31.40 & 32.05 & 32.57 & 32.97 & 33.20 & 33.35 & 33.30 & 32.91 & 33.29 \\
& 40 & 30.39 & 30.55 & 30.46 & 30.05 & 30.70 & 31.39 & 31.75 & 32.06 & 32.15 & 32.21 & 31.77 & 32.18 \\
& 50 & 29.24 & 29.56 & 29.53 & 29.15 & 29.66 & 30.45 & 30.80 & 31.16 & 31.22 & 31.33 & 30.90& 31.33 \\ 
    \bottomrule
  \end{tabular}
  \vspace{-0.35cm}
  \caption{Quantitative comparison on Set8 test set.}
  \label{set8_result}
  \vspace{-0.2cm}
\end{table*}

\section{Further Analysis of Grouped Spatial Shift}
Apart from the local attribute map (LAM) \cite{gu2021interpreting} visualization, we provide further analysis of grouped spatial shift. We perform LAM to obtain the contribution weights of four shifted feature groups of $o_{i+1}$ in helping restoring the local patch of $O_{i}$.  According to the contribution weights, we sort $M$ shift vectors and divide them into different contribution classes. To find the connections between important shifted features and temporal motion information, we select optical flow to evaluate their shift vectors. $M$ shift vectors are sorted according to their contribution weights. We average top-4 important shift vectors obtain a pseudo optical flow $w_{i-1\rightarrow i}$ for the local grids. We also calculate the pseudo optical flows of top $5\sim8$, $9\sim12$ and top $13\sim16$ important vectors. We utilize a pre-trained spynet \cite{Spynet} to estimate the optical flow $W_{i-1\rightarrow i}$ from ground truth clean frames $H_{i-1}$ and $H_{i}$. The optical flow  $W_{i\rightarrow i+1}$ is averaged in the every local grid. We calculate the correlations between optical flow $W_{i-1\rightarrow i}$ and the pseudo optical flow $w_{i-1\rightarrow i}$ along x-axis and y-axis, separately. It is shown in Table~\ref{tb:accuracy_lam} that shifted feature groups usually make more contribution when the shift direction is similar to the optical flow $W_{i-1\rightarrow i}$.
\begin{table}[t]
\centering
\setlength{\tabcolsep}{10pt}
\small
\begin{tabular}{c|cc}
             \hline
                Contribution & $x$-axis & $y$-axis \\ \hline
                Top $1\sim4$ & 71.5 \% & 66.3 \% \\ \hline
                Top $5\sim8$  & 32.5 \% & 34.7 \% \\ \hline
                Top $9\sim12$ & 14.4 \% & 13.8 \% \\ \hline
                Top $13\sim16$ & 3.7 \% & 3.5 \% \\
                \hline
              \end{tabular}
\vspace{-0.1cm}
\caption{The correlations between the optical flow $W_{i-1\rightarrow i}$ of ground truth frames and the pseudo optical flow $w_{i-1\rightarrow i}$ produced from shifted features of different contribution weights.
}
    \label{tb:accuracy_lam}
    \vspace{-0.1cm}
\end{table}

\section{Motion magnitudes} 
We categorize each frame of GoPro dataset according to motion magnitudes. For each blurry frame $I_i$ and its corresponding ground truth $O_i$, we utilize a pre-trained SPyNet to obtain optical flows between $O_i$ and two adjacent frames $O_{i-1}, O_{i+1}$.
We obtain motion magnitudes by averaging the flows. The results in Table ~\ref{tab:motion_magnitude} show that our base model achieves 33.94dB in 10\% largest magnitudes, which surpasses VRT (32.45dB) by +1.49dB. The gain of 10\% smallest is 0.25dB.

\section{Network Architecture}
In our three-stage design, we take a three-scale U-Net \cite{U-Net} as our backbone. For each U-Net, we adopt the U-Net-like structure of MPRNet \cite{Zamir2021MPRNet} to encode effective features. Average pooling and 2D bilinear upsampling is applied to obtain multi-scale features. Each feature in skip connections are processed by a Channel Attention Block (CAB) \cite{zhang2018rcan}, which is the residual blocks equipped with a channel attention layer. The channel attention layer is first introduced in squeeze-and-excitation networks \cite{SENet}, and explored in low-level visions \cite{zhang2018rcan, Zamir2021MPRNet}.
In frame-wise feature extraction and final restoration, We take the Channel Attention Block (CAB) to extract frame-wise features. In multi-frame fusion, we utilize the proposed GSTS blocks to achieve multi-frame feature aggregation and communication. 

\begin{table}[t]
    \small
    \centering
    \begin{tabular}{c|ccc}
    \toprule
    % \hline 
        Method & Largest 10 \% & Smallest 10 \%  & Other 80 \% \\ \hline
        VRT &  32.45  & 35.98 & 34.96 \\ \hline
        Ours & 33.94 $(\textbf{+1.49})$ & 36.23 $(+0.25)$ & 35.61 $(+0.65)$  \\ 
    \bottomrule
    % \hline 
    \end{tabular}
    \vspace{-0.1cm}
    \caption{Deblurring performances of different motion magnitudes.}
    \label{tab:motion_magnitude}
    \vspace{-0.1cm}
\end{table}
A GSTS block contains a grouped spatial-temporal shift operation and a lightweight fusion layer. The fusion layer, consisting of two lightweight convolution blocks (denoted as ``FusionConv''), fuses the spatial-temporal shifted features effectively.
Our FusionConv block takes the framework of Super Kernels (SKFlow) \cite{sun2022skflow}, which utilizes a small kernel convolution and a large kernel convolution as spatial filtering.
The FusionConv block contains three point-wise convolution enable communication across channels and two depth-wise convolution for effective feature fusion. We utilize Layernorm \cite{layernorm} and channel attention \cite{chen2022simple} to improve the network capacity. Learning from NAFNet \cite{chen2022simple}, we replace all GELU layers in SKFLow by gated layers to improve the performance further.

For our small model (``Ours-s''), we stack 3 slim U-Nets with 14 channels for frame-wise processing (Stage-1 and Stage-3) and the channel number of multi-frame fusion is set to be 64. For our base model ``Ours'' and enhanced version ``Ours+'', we stack 5 slim U-Nets with 24 channels for frame-wise processing (Stage-1 and Stage-3) and the channel number of multi-frame fusion is set to be 80. 

%%%%%%%%% REFERENCES
{\small
\bibliographystyle{ieee_fullname}
\bibliography{egbib}

\begin{thebibliography}{10}\itemsep=-1pt

\bibitem{VLNB}
Pablo Arias and Jean-Michel Morel.
\newblock Video denoising via empirical bayesian estimation of space-time
  patches.
\newblock {\em J. Math. Imaging Vis.}, 60(1):70–93, jan 2018.

\bibitem{layernorm}
Jimmy~Lei Ba, Jamie~Ryan Kiros, and Geoffrey~E. Hinton.
\newblock Layer normalization, 2016.

\bibitem{real-timeVSR}
J. Caballero, C. Ledig, A. Aitken, A. Acosta, J. Totz, Z. Wang, and W. Shi.
\newblock Real-time video super-resolution with spatio-temporal networks and
  motion compensation.
\newblock In {\em 2017 IEEE Conference on Computer Vision and Pattern
  Recognition (CVPR)}, pages 2848--2857, Los Alamitos, CA, USA, jul 2017. IEEE
  Computer Society.

\bibitem{chan2021basicvsr}
Kelvin~CK Chan, Xintao Wang, Ke Yu, Chao Dong, and Chen~Change Loy.
\newblock Basicvsr: The search for essential components in video
  super-resolution and beyond.
\newblock In {\em Proceedings of the IEEE conference on computer vision and
  pattern recognition}, 2021.

\bibitem{chan2021understanding}
Kelvin~CK Chan, Xintao Wang, Ke Yu, Chao Dong, and Chen~Change Loy.
\newblock Understanding deformable alignment in video super-resolution.
\newblock In {\em AAAI Conference on Artificial Intelligence}, 2021.

\bibitem{chan2021basicvsr++}
Kelvin~C.K. Chan, Shangchen Zhou, Xiangyu Xu, and Chen~Change Loy.
\newblock {BasicVSR++}: Improving video super-resolution with enhanced
  propagation and alignment.
\newblock In {\em IEEE Conference on Computer Vision and Pattern Recognition},
  2022.

\bibitem{chan2022investigating}
Kelvin~C.K. Chan, Shangchen Zhou, Xiangyu Xu, and Chen~Change Loy.
\newblock Investigating tradeoffs in real-world video super-resolution.
\newblock In {\em IEEE Conference on Computer Vision and Pattern Recognition},
  2022.

\bibitem{chan2022generalization}
Kelvin~CK Chan, Shangchen Zhou, Xiangyu Xu, and Chen~Change Loy.
\newblock On the generalization of {BasicVSR++} to video deblurring and
  denoising.
\newblock {\em arXiv preprint arXiv:2204.05308}, 2022.

\bibitem{RNN-MBP}
Zhu Chao, Dong Hang, Pan Jinshan, Liang Boyang, Huang Yuhao, Fu Lean, and Wang
  Fei.
\newblock Deep recurrent neural network with multi-scale bi-directional
  propagation for video deblurring.
\newblock In {\em AAAI}, 2022.

\bibitem{SeeMotion}
Chen Chen, Qifeng Chen, Minh~N. Do, and Vladlen Koltun.
\newblock Seeing motion in the dark.
\newblock In {\em Proceedings of the IEEE/CVF International Conference on
  Computer Vision (ICCV)}, October 2019.

\bibitem{chen2022simple}
Liangyu Chen, Xiaojie Chu, Xiangyu Zhang, and Jian Sun.
\newblock Simple baselines for image restoration.
\newblock {\em arXiv preprint arXiv:2204.04676}, 2022.

\bibitem{shifts_conv_need}
W. Chen, D. Xie, Y. Zhang, and S. Pu.
\newblock All you need is a few shifts: Designing efficient convolutional
  neural networks for image classification.
\newblock In {\em 2019 IEEE/CVF Conference on Computer Vision and Pattern
  Recognition (CVPR)}, pages 7234--7243, Los Alamitos, CA, USA, jun 2019. IEEE
  Computer Society.

\bibitem{dai2017deformable}
Jifeng Dai, Haozhi Qi, Yuwen Xiong, Yi Li, Guodong Zhang, Han Hu, and Yichen
  Wei.
\newblock Deformable convolutional networks.
\newblock In {\em Proceedings of the IEEE international conference on computer
  vision}, 2017.

\bibitem{devlin2018bert}
Jacob Devlin, Ming-Wei Chang, Kenton Lee, and Kristina Toutanova.
\newblock Bert: Pre-training of deep bidirectional transformers for language
  understanding.
\newblock {\em arXiv preprint arXiv:1810.04805}, 2018.

\bibitem{ding2021repvgg}
Xiaohan Ding, Xiangyu Zhang, Ningning Ma, Jungong Han, Guiguang Ding, and Jian
  Sun.
\newblock Repvgg: Making vgg-style convnets great again.
\newblock In {\em Proceedings of the IEEE/CVF Conference on Computer Vision and
  Pattern Recognition}, pages 13733--13742, 2021.

\bibitem{replknet}
Xiaohan Ding, Xiangyu Zhang, Yizhuang Zhou, Jungong Han, Guiguang Ding, and
  Jian Sun.
\newblock Scaling up your kernels to 31x31: Revisiting large kernel design in
  cnns.
\newblock {\em arXiv preprint arXiv:2203.06717}, 2022.

\bibitem{dosovitskiy2021an}
Alexey Dosovitskiy, Lucas Beyer, Alexander Kolesnikov, Dirk Weissenborn,
  Xiaohua Zhai, Thomas Unterthiner, Mostafa Dehghani, Matthias Minderer, Georg
  Heigold, Sylvain Gelly, Jakob Uszkoreit, and Neil Houlsby.
\newblock An image is worth 16x16 words: Transformers for image recognition at
  scale.
\newblock In {\em International Conference on Learning Representations}, 2021.

\bibitem{gu2021interpreting}
Jinjin Gu and Chao Dong.
\newblock Interpreting super-resolution networks with local attribution maps.
\newblock In {\em Proceedings of the IEEE/CVF Conference on Computer Vision and
  Pattern Recognition}, pages 9199--9208, 2021.

\bibitem{Resnet}
Kaiming He, Xiangyu Zhang, Shaoqing Ren, and Jian Sun.
\newblock Deep residual learning for image recognition.
\newblock In {\em The IEEE Conference on Computer Vision and Pattern
  Recognition (CVPR)}, June 2016.

\bibitem{SENet}
Jie Hu, Li Shen, and Gang Sun.
\newblock Squeeze-and-excitation networks.
\newblock In {\em 2018 IEEE/CVF Conference on Computer Vision and Pattern
  Recognition}, pages 7132--7141, 2018.

\bibitem{Compression2022}
Cong Huang, Jiahao Li, Bin Li, Dong Liu, and Yan Lu.
\newblock Neural compression-based feature learning for video restoration,
  2022.

\bibitem{huang2022flowformer}
Zhaoyang Huang, Xiaoyu Shi, Chao Zhang, Qiang Wang, Ka~Chun Cheung, Hongwei
  Qin, Jifeng Dai, and Hongsheng Li.
\newblock Flowformer: A transformer architecture for optical flow.
\newblock {\em arXiv preprint arXiv:2203.16194}, 2022.

\bibitem{RSDN}
Takashi Isobe, Xu Jia, Shuhang Gu, Songjiang Li, Shengjin Wang, and Qi Tian.
\newblock Video super-resolution with recurrent structure-detail network.
\newblock {\em CoRR}, abs/2008.00455, 2020.

\bibitem{Shift_conv_fast}
Yunho Jeon and Junmo Kim.
\newblock Constructing fast network through deconstruction of convolution.
\newblock In {\em Proceedings of the 32nd International Conference on Neural
  Information Processing Systems}, NIPS'18, page 5955–5965, Red Hook, NY,
  USA, 2018. Curran Associates Inc.

\bibitem{ERDN}
Bangrui Jiang, Zhihuai Xie, Zhen Xia, Songnan Li, and Shan Liu.
\newblock Erdn: Equivalent receptive field deformable network for video
  deblurring.
\newblock In Shai Avidan, Gabriel Brostow, Moustapha Ciss{\'e}, Giovanni~Maria
  Farinella, and Tal Hassner, editors, {\em Computer Vision -- ECCV 2022},
  pages 663--678, Cham, 2022. Springer Nature Switzerland.

\bibitem{Jo2018dynamic_upsampling}
Younghyun Jo, Seoung~Wug Oh, Jaeyeon Kang, and Seon~Joo Kim.
\newblock Deep video super-resolution network using dynamic upsampling filters
  without explicit motion compensation.
\newblock In {\em 2018 IEEE/CVF Conference on Computer Vision and Pattern
  Recognition}, pages 3224--3232, 2018.

\bibitem{VSRCNN}
Armin Kappeler, Seunghwan Yoo, Qiqin Dai, and Aggelos~K. Katsaggelos.
\newblock Video super-resolution with convolutional neural networks.
\newblock {\em IEEE Transactions on Computational Imaging}, 2(2):109--122,
  2016.

\bibitem{khoreva2018video}
Anna Khoreva, Anna Rohrbach, and Bernt Schiele.
\newblock Video object segmentation with language referring expressions.
\newblock In {\em Asian Conference on Computer Vision}, pages 123--141.
  Springer, 2018.

\bibitem{adam}
Diederik~P. Kingma and Jimmy Ba.
\newblock Adam: {A} method for stochastic optimization.
\newblock In Yoshua Bengio and Yann LeCun, editors, {\em 3rd International
  Conference on Learning Representations, {ICLR} 2015, San Diego, CA, USA, May
  7-9, 2015, Conference Track Proceedings}, 2015.

\bibitem{ARVo}
Dongxu Li, Chenchen Xu, Kaihao Zhang, Xin~Yu 0002, Yiran Zhong, Wenqi Ren,
  Hanna Suominen, and Hongdong Li.
\newblock Arvo: Learning all-range volumetric correspondence for video
  deblurring.
\newblock In {\em IEEE Conference on Computer Vision and Pattern Recognition,
  CVPR 2021, virtual, June 19-25, 2021}, pages 7721--7731. Computer Vision
  Foundation / IEEE, 2021.

\bibitem{Li2022Learning}
Dasong Li, Yi Zhang, Ka~Chun Cheung, Xiaogang Wang, Hongwei Qin, and Hongsheng
  Li.
\newblock Learning degradation representations for image deblurring.
\newblock In Shai Avidan, Gabriel Brostow, Moustapha Ciss{\'e}, Giovanni~Maria
  Farinella, and Tal Hassner, editors, {\em Computer Vision -- ECCV 2022},
  pages 736--753, Cham, 2022. Springer Nature Switzerland.

\bibitem{Li2022efficient}
Dasong Li, Yi Zhang, Ka~Lung Law, Xiaogang Wang, Hongwei Qin, and Hongsheng Li.
\newblock Efficient burst raw denoising with variance stabilization and
  multi-frequency denoising network, 2022.

\bibitem{li2022unidirectional}
Junyi Li, Xiaohe Wu, Zhenxing Niu, and Wangmeng Zuo.
\newblock Unidirectional video denoising by mimicking backward recurrent
  modules with look-ahead forward ones.
\newblock {\em arXiv preprint arXiv:2204.05532}, 2022.

\bibitem{AShift-MLP}
Dongze Lian, Zehao Yu, Xing Sun, and Shenghua Gao.
\newblock As-mlp: An axial shifted mlp architecture for vision, 2021.

\bibitem{liang2022vrt}
Jingyun Liang, Jiezhang Cao, Yuchen Fan, Kai Zhang, Rakesh Ranjan, Yawei Li,
  Radu Timofte, and Luc Van~Gool.
\newblock Vrt: A video restoration transformer.
\newblock {\em arXiv preprint arXiv:2201.12288}, 2022.

\bibitem{liang2022rvrt}
Jingyun Liang, Yuchen Fan, Xiaoyu Xiang, Rakesh Ranjan, Eddy Ilg, Simon Green,
  Jiezhang Cao, Kai Zhang, Radu Timofte, and Luc Van~Gool.
\newblock Recurrent video restoration transformer with guided deformable
  attention.
\newblock {\em arXiv preprint arXiv:2206.02146}, 2022.

\bibitem{lin2019tsm}
Ji Lin, Chuang Gan, and Song Han.
\newblock Tsm: Temporal shift module for efficient video understanding.
\newblock In {\em Proceedings of the IEEE International Conference on Computer
  Vision}, 2019.

\bibitem{liu2021Swin}
Ze Liu, Yutong Lin, Yue Cao, Han Hu, Yixuan Wei, Zheng Zhang, Stephen Lin, and
  Baining Guo.
\newblock Swin transformer: Hierarchical vision transformer using shifted
  windows.
\newblock In {\em Proceedings of the IEEE/CVF International Conference on
  Computer Vision (ICCV)}, 2021.

\bibitem{convNet2020}
Zhuang Liu, Hanzi Mao, Chao{-}Yuan Wu, Christoph Feichtenhofer, Trevor Darrell,
  and Saining Xie.
\newblock A convnet for the 2020s.
\newblock {\em CoRR}, abs/2201.03545, 2022.

\bibitem{cosine_annealing}
Ilya Loshchilov and Frank Hutter.
\newblock {SGDR:} stochastic gradient descent with warm restarts.
\newblock In {\em 5th International Conference on Learning Representations,
  {ICLR} 2017, Toulon, France, April 24-26, 2017, Conference Track
  Proceedings}. OpenReview.net, 2017.

\bibitem{luo2017understanding}
Wenjie Luo, Yujia Li, Raquel Urtasun, and Richard Zemel.
\newblock Understanding the effective receptive field in deep convolutional
  neural networks.
\newblock In {\em Proceedings of the 30th International Conference on Neural
  Information Processing Systems}, NIPS'16, page 4905–4913, Red Hook, NY,
  USA, 2016. Curran Associates Inc.

\bibitem{EMVD}
M. Maggioni, Y. Huang, C. Li, S. Xiao, Z. Fu, and F. Song.
\newblock Efficient multi-stage video denoising with recurrent spatio-temporal
  fusion.
\newblock In {\em 2021 IEEE/CVF Conference on Computer Vision and Pattern
  Recognition (CVPR)}, 2021.

\bibitem{MildenhallKPN18}
Ben Mildenhall, Jonathan~T. Barron, Jiawen Chen, Dillon Sharlet, Ren Ng, and
  Robert Carroll.
\newblock Burst denoising with kernel prediction networks.
\newblock In {\em {CVPR}}, 2018.

\bibitem{reds4}
Seungjun Nah, Sungyong Baik, Seokil Hong, Gyeongsik Moon, Sanghyun Son, Radu
  Timofte, and Kyoung~Mu Lee.
\newblock Ntire 2019 challenge on video deblurring and super-resolution:
  Dataset and study.
\newblock In {\em 2019 IEEE/CVF Conference on Computer Vision and Pattern
  Recognition Workshops (CVPRW)}, pages 1996--2005, 2019.

\bibitem{deblur-multi-scale}
Seungjun Nah, Tae~Hyun Kim, and Kyoung~Mu Lee.
\newblock Deep multi-scale convolutional neural network for dynamic scene
  deblurring.
\newblock In {\em The IEEE Conference on Computer Vision and Pattern
  Recognition (CVPR)}, July 2017.

\bibitem{hourglass2016}
Alejandro Newell, Kaiyu Yang, and Jia Deng.
\newblock Stacked hourglass networks for human pose estimation.
\newblock In Bastian Leibe, Jiri Matas, Nicu Sebe, and Max Welling, editors,
  {\em Computer Vision -- ECCV 2016}, pages 483--499, Cham, 2016. Springer
  International Publishing.

\bibitem{Pan2020cdvdtsp}
Jinshan Pan, Haoran Bai, and Jinhui Tang.
\newblock Cascaded deep video deblurring using temporal sharpness prior.
\newblock In {\em IEEE/CVF Conference on Computer Vision and Pattern
  Recognition (CVPR)}, June 2020.

\bibitem{Spynet}
Anurag Ranjan and Michael~J. Black.
\newblock Optical flow estimation using a spatial pyramid network.
\newblock In {\em 2017 IEEE Conference on Computer Vision and Pattern
  Recognition (CVPR)}, pages 2720--2729, 2017.

\bibitem{tsmdenoising}
Xuejian Rong, Denis Demandolx, Kevin Matzen, Priyam Chatterjee, and Yingli
  Tian.
\newblock Burst denoising via temporally shifted wavelet transforms.
\newblock In {\em Computer Vision – ECCV 2020: 16th European Conference,
  Glasgow, UK, August 23–28, 2020, Proceedings, Part XIII}, page 240–256,
  Berlin, Heidelberg, 2020. Springer-Verlag.

\bibitem{U-Net}
Olaf Ronneberger, Philipp Fischer, and Thomas Brox.
\newblock U-net: Convolutional networks for biomedical image segmentation.
\newblock In Nassir Navab, Joachim Hornegger, William~M. Wells, and
  Alejandro~F. Frangi, editors, {\em Medical Image Computing and
  Computer-Assisted Intervention -- MICCAI 2015}, 2015.

\bibitem{shi2023videoflow}
Xiaoyu Shi, Zhaoyang Huang, Weikang Bian, Dasong Li, Manyuan Zhang, Ka~Chun
  Cheung, Simon See, Hongwei Qin, Jifeng Dai, and Hongsheng Li.
\newblock Videoflow: Exploiting temporal cues for multi-frame optical flow
  estimation.
\newblock {\em arXiv preprint arXiv:2303.08340}, 2023.

\bibitem{shi2023flowformer++}
Xiaoyu Shi, Zhaoyang Huang, Dasong Li, Manyuan Zhang, Ka~Chun Cheung, Simon
  See, Hongwei Qin, Jifeng Dai, and Hongsheng Li.
\newblock Flowformer++: Masked cost volume autoencoding for pretraining optical
  flow estimation.
\newblock {\em arXiv preprint arXiv:2303.01237}, 2023.

\bibitem{Son2021PVDNet}
Hyeongseok Son, Junyong Lee, Jonghyeop Lee, Sunghyun Cho, and Seungyong Lee.
\newblock Recurrent video deblurring with blur-invariant motion estimation and
  pixel volumes.
\newblock {\em ACM Transactions on Graphics (TOG)}, 40(5), 2021.

\bibitem{tempformer}
Mingyang Song, Yang Zhang, and Tun\c{c}~O. Ayd\i{}n.
\newblock Tempformer: Temporally consistent transformer for video denoising.
\newblock In {\em Computer Vision – ECCV 2022: 17th European Conference, Tel
  Aviv, Israel, October 23–27, 2022, Proceedings, Part XIX}, 2022.

\bibitem{su2017deblurring}
Shuochen Su, Mauricio Delbracio, Jue Wang, Guillermo Sapiro, Wolfgang Heidrich,
  and Oliver Wang.
\newblock Deep video deblurring for hand-held cameras.
\newblock In {\em 2017 IEEE Conference on Computer Vision and Pattern
  Recognition (CVPR)}, pages 237--246, 2017.

\bibitem{sun2022skflow}
SHANGKUN SUN, Yuanqi Chen, Yu Zhu, Guodong Guo, and Ge Li.
\newblock {SKF}low: Learning optical flow with super kernels.
\newblock In Alice~H. Oh, Alekh Agarwal, Danielle Belgrave, and Kyunghyun Cho,
  editors, {\em Advances in Neural Information Processing Systems}, 2022.

\bibitem{dvdnet}
Matias Tassano, Julie Delon, and Thomas Veit.
\newblock {dvdnet: a fast network for deep video denoising}.
\newblock In {\em {2019 IEEE International Conference on Image Processing}},
  Taipei, Taiwan, Sept. 2019.

\bibitem{fastdvdnet}
Matias Tassano, Julie Delon, and Thomas Veit.
\newblock Fastdvdnet: Towards real-time deep video denoising without flow
  estimation.
\newblock In {\em Proceedings of the IEEE/CVF Conference on Computer Vision and
  Pattern Recognition (CVPR)}, June 2020.

\bibitem{tian2020tdan}
Yapeng Tian, Yulun Zhang, Yun Fu, and Chenliang Xu.
\newblock Tdan: Temporally-deformable alignment network for video
  super-resolution.
\newblock In {\em The IEEE Conference on Computer Vision and Pattern
  Recognition (CVPR)}, June 2020.

\bibitem{Vaksman2021PatchC}
G. Vaksman, M. Elad, and P. Milanfar.
\newblock Patch craft: Video denoising by deep modeling and patch matching.
\newblock In {\em 2021 IEEE/CVF International Conference on Computer Vision
  (ICCV)}, pages 2137--2146, Los Alamitos, CA, USA, oct 2021. IEEE Computer
  Society.

\bibitem{msra_shift}
Guangting Wang, Yucheng Zhao, Chuanxin Tang, Chong Luo, and Wenjun Zeng.
\newblock When shift operation meets vision transformer: An extremely simple
  alternative to attention mechanism, 2022.

\bibitem{wang2019edvr}
Xintao Wang, Kelvin~C.K. Chan, Ke Yu, Chao Dong, and Chen~Change Loy.
\newblock Edvr: Video restoration with enhanced deformable convolutional
  networks.
\newblock In {\em The IEEE Conference on Computer Vision and Pattern
  Recognition (CVPR) Workshops}, June 2019.

\bibitem{deepflow}
Philippe Weinzaepfel, Jerome Revaud, Zaid Harchaoui, and Cordelia Schmid.
\newblock Deepflow: Large displacement optical flow with deep matching.
\newblock In {\em 2013 IEEE International Conference on Computer Vision}, pages
  1385--1392, 2013.

\bibitem{wu2018shiftconv}
B. Wu, A. Wan, X. Yue, P. Jin, S. Zhao, N. Golmant, A. Gholaminejad, J.
  Gonzalez, and K. Keutzer.
\newblock Shift: A zero flop, zero parameter alternative to spatial
  convolutions.
\newblock In {\em 2018 IEEE/CVF Conference on Computer Vision and Pattern
  Recognition (CVPR)}, pages 9127--9135, Los Alamitos, CA, USA, jun 2018. IEEE
  Computer Society.

\bibitem{BPN}
Z. Xia, F. Perazzi, M. Gharbi, K. Sunkavalli, and A. Chakrabarti.
\newblock Basis prediction networks for effective burst denoising with large
  kernels.
\newblock In {\em 2020 IEEE/CVF Conference on Computer Vision and Pattern
  Recognition (CVPR)}, 2020.

\bibitem{toflow}
Tianfan Xue, Baian Chen, Jiajun Wu, Donglai Wei, and William~T Freeman.
\newblock Video enhancement with task-oriented flow.
\newblock {\em International Journal of Computer Vision (IJCV)},
  127(8):1106--1125, 2019.

\bibitem{S2-MLP}
T. Yu, X. Li, Y. Cai, M. Sun, and P. Li.
\newblock S2-mlp: Spatial-shift mlp architecture for vision.
\newblock In {\em 2022 IEEE/CVF Winter Conference on Applications of Computer
  Vision (WACV)}, pages 3615--3624, Los Alamitos, CA, USA, jan 2022. IEEE
  Computer Society.

\bibitem{RViDeNet}
Huanjing Yue, Cong Cao, Lei Liao, Ronghe Chu, and Jingyu Yang.
\newblock Supervised raw video denoising with a benchmark dataset on dynamic
  scenes.
\newblock In {\em IEEE Conference on Computer Vision and Pattern Recognition},
  2020.

\bibitem{Zamir2021MPRNet}
Syed~Waqas Zamir, Aditya Arora, Salman Khan, Munawar Hayat, Fahad~Shahbaz Khan,
  Ming-Hsuan Yang, and Ling Shao.
\newblock Multi-stage progressive image restoration.
\newblock In {\em CVPR}, 2021.

\bibitem{zhang2022spatio}
Huicong Zhang, Haozhe Xie, and Hongxun Yao.
\newblock Spatio-temporal deformable attention network for video deblurring.
\newblock In {\em ECCV}, 2022.

\bibitem{zhang2022efficient}
Xindong Zhang, Hui Zeng, Shi Guo, and Lei Zhang.
\newblock Efficient long-range attention network for image super-resolution.
\newblock In {\em European Conference on Computer Vision}, 2022.

\bibitem{zhang2023kbnet}
Yi Zhang, Dasong Li, Xiaoyu Shi, Dailan He, Kangning Song, Xiaogang Wang,
  Hongwei Qin, and Hongsheng Li.
\newblock Kbnet: Kernel basis network for image restoration, 2023.

\bibitem{zhang2018rcan}
Yulun Zhang, Kunpeng Li, Kai Li, Lichen Wang, Bineng Zhong, and Yun Fu.
\newblock Image super-resolution using very deep residual channel attention
  networks.
\newblock In {\em ECCV}, 2018.

\bibitem{zhou2019stfan}
Shangchen Zhou, Jiawei Zhang, Jinshan Pan, Haozhe Xie, Wangmeng Zuo, and Jimmy
  Ren.
\newblock Spatio-temporal filter adaptive network for video deblurring.
\newblock In {\em Proceedings of the IEEE International Conference on Computer
  Vision}, 2019.

\bibitem{zhu2019deformable}
Xizhou Zhu, Han Hu, Stephen Lin, and Jifeng Dai.
\newblock Deformable convnets v2: More deformable, better results.
\newblock In {\em Proceedings of the IEEE/CVF Conference on Computer Vision and
  Pattern Recognition}, pages 9308--9316, 2019.

\end{thebibliography}
}

\end{document}